\documentclass[fleqn,usenatbib]{mnras}
\usepackage[dvipsnames]{xcolor}

\usepackage{newtxtext,newtxmath}
\usepackage{mdframed}
\usepackage{graphicx}
\usepackage{subcaption}
\usepackage{dsfont}
\usepackage{booktabs}

\usepackage{hyperref}
\usepackage[T1]{fontenc}
\usepackage{ae,aecompl}
\usepackage{todonotes}
\usepackage{multirow}
\usepackage{booktabs}
\usepackage[dvipsnames]{xcolor}

\usepackage{color}

\newcommand{\blue}[1]{\textcolor{blue}{#1}}

\usepackage{comment}
\usepackage{listings}

\lstdefinestyle{shadedcode}{
    backgroundcolor=\color{gray!10},  
    frame=single,                     
    rulecolor=\color{gray!50},        
    basicstyle=\tiny\ttfamily,  
    breaklines=true,                  
    captionpos=b,                     
    keepspaces=true,                  
    showspaces=false,                 
    showstringspaces=false,           
    showtabs=false,                   
    tabsize=4,                        
    xleftmargin=\parindent,           
    xrightmargin=0.3cm,               
    framexleftmargin=6pt,             
    framexrightmargin=6pt,            
    framextopmargin=3pt,              
    framexbottommargin=3pt,           
}

\usepackage{lineno}

\author[Prat et al.]{
\parbox{\textwidth}{
{\fontsize{11.3}{12} \selectfont 
J.~Prat,$^{1,2,3}$\thanks{E-mail:judit.prat@su.se},
M.~Gatti,$^{4}$
C.~Doux,$^{5,6}$
P.~Pranav,$^{7,8}$
C.~Chang,$^{3,4}$
N.~Jeffrey,$^{9}$
L.~Whiteway,$^{9}$
D.~Anbajagane,$^{4}$
S.~Sugiyama,$^{5}$
A.~Thomsen,$^{10}$
A.~Alarcon,$^{11}$
A.~Amon,$^{12}$
K.~Bechtol,$^{13}$
G.~M.~Bernstein,$^{5}$
A.~Campos,$^{14,15}$
R.~Chen,$^{16}$
A.~Choi,$^{17}$
C.~Davis,$^{18}$
J.~DeRose,$^{19}$
S.~Dodelson,$^{3,20,4}$
K.~Eckert,$^{5}$
J.~Elvin-Poole,$^{21}$
S.~Everett,$^{22}$
A.~Fert\'e,$^{23}$
D.~Gruen,$^{24}$
E.~M.~Huff,$^{25}$
I.~Harrison,$^{26}$
K.~Herner,$^{20}$
M.~Jarvis,$^{5}$
N.~Kuropatkin,$^{20}$
P.-F.~Leget,$^{18}$
N.~MacCrann,$^{27}$
J.~McCullough,$^{12,18,23,24}$
J.~Myles,$^{12}$
A. Navarro-Alsina,$^{28}$
S.~Pandey,$^{5}$
M.~Raveri,$^{29}$
R.~P.~Rollins,$^{30}$
A.~Roodman,$^{23,18}$
C.~S{\'a}nchez,$^{5}$
L.~F.~Secco,$^{4}$
E.~Sheldon,$^{31}$
T.~Shin,$^{32}$
M.~A.~Troxel,$^{16}$
I.~Tutusaus,$^{33}$
T.~N.~Varga,$^{34,35,36}$
B.~Yanny,$^{20}$
B.~Yin,$^{14}$
Y.~Zhang,$^{37}$
J.~Zuntz,$^{38}$
T.~M.~C.~Abbott,$^{37}$
M.~Aguena,$^{39,40}$
S.~Allam,$^{20}$
F.~Andrade-Oliveira,$^{41}$
J.~Blazek,$^{42}$
S.~Bocquet,$^{24}$
D.~Brooks,$^{9}$
J.~Carretero,$^{43}$
A.~Carnero~Rosell,$^{44,40,45}$
R.~Cawthon,$^{46}$
J.~De~Vicente,$^{47}$
S.~Desai,$^{48}$
M.~E.~da Silva Pereira,$^{49}$
H.~T.~Diehl,$^{20}$
B.~Flaugher,$^{20}$
J.~Frieman,$^{3,20,4}$
J.~Garc\'ia-Bellido,$^{50}$
R.~A.~Gruendl,$^{51,52}$
G.~Gutierrez,$^{20}$
S.~R.~Hinton,$^{53}$
D.~L.~Hollowood,$^{54}$
K.~Honscheid,$^{55,56}$
D.~J.~James,$^{57}$
K.~Kuehn,$^{58,59}$
L.~N.~da Costa,$^{40}$
O.~Lahav,$^{9}$
S.~Lee,$^{25}$
J.~L.~Marshall,$^{60}$
J. Mena-Fern{\'a}ndez,$^{6}$
R.~Miquel,$^{61,43}$
J.~J.~Mohr,$^{24}$
R.~L.~C.~Ogando,$^{62,63}$
A.~A.~Plazas~Malag\'on,$^{18,23}$
A.~Porredon,$^{47,64}$
S.~Samuroff,$^{42,43}$
E.~Sanchez,$^{47}$
B.~Santiago,$^{65,40}$
I.~Sevilla-Noarbe,$^{47}$
M.~Smith,$^{66}$
E.~Suchyta,$^{67}$
M.~E.~C.~Swanson,$^{51}$
D.~Thomas,$^{68}$
C.~To,$^{3}$
V.~Vikram,$^{69}$
A.~R.~Walker,$^{37}$
N.~Weaverdyck,$^{70,19}$
and J.~Weller$^{35,36}$
\begin{center} (The DES Collaboration) \end{center}
}}
\vspace{0.4cm}
\\
\parbox{\textwidth}{Author affiliations are listed at the end of the paper}}

\usepackage{eso-pic}

\AddToShipoutPictureBG*{%
  \AtPageUpperLeft{%
    \hspace{0.75\paperwidth}%
    \raisebox{-4.5\baselineskip}{%
      \makebox[0pt][l]{\textnormal{DES-2025-0898}}
}}}%

\AddToShipoutPictureBG*{%
  \AtPageUpperLeft{%
    \hspace{0.75\paperwidth}%
    \raisebox{-5.5\baselineskip}{%
      \makebox[0pt][l]{\textnormal{FERMILAB-PUB-25-0327-PPD}}
}}}%

\title[Persistent homology on weak lensing maps]{Dark Energy Survey Year 3 results: $w$CDM cosmology from simulation-based inference with persistent homology on the sphere}

\begin{document}
\label{firstpage}
\pagerange{\pageref{firstpage}--\pageref{lastpage}}
\maketitle

\begin{abstract}
We present cosmological constraints from Dark Energy Survey Year 3 (DES Y3) weak lensing data using persistent homology, a topological data analysis technique that tracks how features like clusters and voids evolve across density thresholds. For the first time, we apply spherical persistent homology to galaxy survey data through the algorithm \textsc{TopoS2}, which is optimized for curved-sky analyses and \textsc{HEALPix} compatibility. Employing a simulation-based inference framework with the Gower Street simulation suite---specifically designed to mimic DES Y3 data properties---we extract topological summary statistics from  convergence maps across multiple smoothing scales and redshift bins. After neural network compression of these statistics, we estimate the likelihood function and validate our analysis against baryonic feedback effects, finding minimal biases (under $0.3\sigma$) in the $\Omega_\mathrm{m}-S_8$ plane. Assuming the $w$CDM model, our combined Betti numbers and second moments analysis yields $S_8 = 0.821 \pm 0.018$ and $\Omega_\mathrm{m} = 0.304\pm0.037$---constraints 70\% tighter than those from cosmic shear two-point statistics in the same parameter plane. Our results demonstrate that topological methods provide a powerful and robust framework for extracting cosmological information, with our spherical methodology readily applicable to upcoming Stage IV wide-field galaxy surveys.
\end{abstract}
\begin{keywords} 
cosmology: observations -- cosmological parameters -- gravitational lensing: weak --  large-scale structure of Universe
\end{keywords}

\section{Introduction} \label{sec:intro}

The current era of precision cosmology demands maximizing information extraction from cosmic large-scale structure while maintaining  robustness. Several approaches address this challenge with varying trade-offs. Full-field inference methods \citep{Lanzieri2024, Zhou2025} theoretically capture maximal information, but their practical application to observational datasets remains difficult despite recent advancements \citep{Porqueres2023}. Conversely, the standard 3$\times$2pt methodology--combining three two-point correlation functions from weak lensing and galaxy clustering--has successfully constrained cosmology in Stage III photometric surveys \citep{Heymans2021, Abbott2022, Sugiyama2023}, but inherently misses substantial information. This limitation arises because two-point statistics only capture Gaussian features, while the late-time Universe is significantly non-Gaussian. The middle-ground approach using \textit{higher-order} summary statistics offers a promising compromise, balancing a practical implementation with the ability to extract substantially more cosmological information than traditional methods.

Indeed, numerous higher-order statistics have already been successfully applied to extract cosmological information from real galaxy survey data, particularly from weak lensing observations. These include third-order statistics applied to both the Dark Energy Survey (DES) \citep{Gatti2022, Gatti2025} and the Kilo-Degree Survey (KiDS) \citep{Burger2024}; scattering transforms and wavelet phase harmonics implemented on DES \citep{Gatti2025}, and scattering transforms on the Hyper Suprime-Cam survey (HSC) \citep{Cheng2025}; Minkowski functionals, peak and minima counts, and convergence probability distribution functions (PDFs) analyzed in HSC data \citep{Novaes2025}; CNN-based and peak count analyses in DES \citep{Jeffrey2025}; Minkowski functionals on HSC \citep{Armijo2025}; joint DES and KiDS analysis of peak and minima counts \citep{Harnois2024}; and density-split statistics applied to DES \citep{Gruen2018}  and to KiDS \citep{Burger2023}.

When inferring cosmological information from higher-order statistics one often encounters a significant challenge: the absence of analytical models. However, the landscape of this research field has dramatically shifted in recent years, thanks to the advent of simulation-based inference (SBI). This approach is based on the creation of large synthetic data sets that closely replicate the features of actual astronomical observations and that are generated at different input target parameters. Then, a given summary statistic is measured in each simulation, thereby providing a \textit{model} for such statistic. SBI relies on the generation of realistic enough simulations, which need to mirror the characteristics of real observations. In this work, we used the Gower St. Simulations suite, released in \cite{Jeffrey2025}. This publicly available simulation suite\footnote{\url{http://www.star.ucl.ac.uk/GowerStreetSims/}} was designed for DES inference, with additional DES Y3 validation presented in \cite{Gatti2024}.

Interestingly, the SBI approach  not only addresses the challenge of modeling higher-order statistics without analytical models but also has a multitude of additional benefits. Notably, it eliminates the need to predefine a likelihood form; instead, the likelihood is \textit{learned} from the simulations, lending this method the alternative name of likelihood-free inference. This capability allows for the empirical validation of commonly assumed Gaussian likelihood forms in standard analyses. The covariance matrix for the statistics is also not needed. Another major advantage is the method's efficiency in handling high-dimensional nuisance parameters that we often want to marginalize over, such as uncertainties in the redshift distributions. While recently developed approaches to marginalize over the full shape of the redshift distributions (such as \textsc{Hyperrank}, \citealt{Cordero2022}) turned out to be too computationally burdensome, in the SBI framework nuisance parameter uncertainties are  straightforward to incorporate. They are simply varied within the input simulations following the desired prior distribution, making the marginalization process effortless. Lastly, once the likelihood is learned through this method, sampling it with standard techniques such as Markov Chain Monte Carlo (MCMC) becomes orders of magnitude less computationally expensive compared to using analytical models. 

In this work we apply persistent homology to curved-sky weak lensing mass maps generated from the first three years of observations from the Dark Energy Survey (DES Y3), described in detail in  \citet*{Jeffrey2021}.  Persistent homology, and more broadly the field of topological data analysis (TDA), seeks to quantify the shape and structure of data across different scales. It has a broad range of applications: within cosmology only it has been used to probe topological structures of the cosmic web as in e.g. \citet{Wilding2021, Sousbie2011, Tsizh2023, Elbers2023}, primordial non-Gaussianities \citep{Cole2020}, small-scale subhalo distributions \citep{Cisewski-Kehe2022}, the impact of massive neutrinos on the large-scale structure \citep{Jalali2023}, extracting cosmological information \citep{Yip2024, Calles2024, Yip2025} and others. Persistent homology has also been used before to extract higher order information from weak lensing maps  in \citet{Heydenreich2022}, where it was applied to the DES Y1 data set. Persistent homology is closely related to other higher order statistics commonly applied to weak lensing mass maps. Notably, certain Minkowski functionals (MFs)—such as genus and Euler characteristic—are already encapsulated within persistent homology statistics. Beyond incorporating these MF elements, persistent homology includes peaks and minima counts. 

In this work we apply persistent homology to the DES Y3 data set with two significant methodological advancements over \citet{Heydenreich2022}. First, we employ \textsc{TopoS2}, an algorithm specifically designed for spherical geometry and to interface with the \textsc{HEALPix} pixelization scheme. Originally developed for Cosmic Microwave Background analysis \citep{Pranav2022}, this work represents the first use of \textit{spherical} persistent homology in galaxy survey data. This advancement --allowing analysis of complete surveys in their native spherical geometry rather than requiring fragmentation and flat-plane projection-- will become increasingly valuable for upcoming large-area surveys like the Vera C. Rubin Observatory's Legacy Survey of Space and Time (\citealt{abell2009lsst}) and Euclid \citep{Euclid}.

Secondly, we integrate a robust, validated simulation-based inference framework that is consistent with two other companion analyses that apply other  higher-order summary statistic to DES Y3 data: Peaks and Convolutional Neural Networks (CNN) and angular power spectra in \citet{Jeffrey2025}, and wavelet harmonics, scattering transforms and moments in \citet{Gatti2024,Gatti2025}. This cohesive approach facilitates a straightforward comparison of the constraining power, performance, and dependence on systematics among various summary statistics. While \citet{EuclidCollaboration2023} pioneered forecasting the relative power of ten distinct higher-order weak lensing statistics using Fisher matrices and simplifying assumptions, our work provides similar comparative insights using real observational data--offering valuable guidance for applications to future datasets.

This paper is structured as follows: in Sec.~\ref{sec:sbi} we present the simulation-based inference framework; in Sec.~\ref{sec:theory} we summarize the weak gravitational lensing theory relevant for this work; in Sec.~\ref{sec:data}  we describe our data and simulations; in Sec.~\ref{sec:persistent-homology} we present the persistent homology statistic and describe how we apply it to weak lensing mass maps; in Sec.~\ref{sec:compression} we describe the compression models, in Sec.~\ref{sec:validation} we include the validation tests;  in Sec.~\ref{sec:cosmo-results} we present the cosmological results and we conclude in Sec.~\ref{sec:conclusion}.

\begin{figure*}
   \centering
    \includegraphics[width=\textwidth]{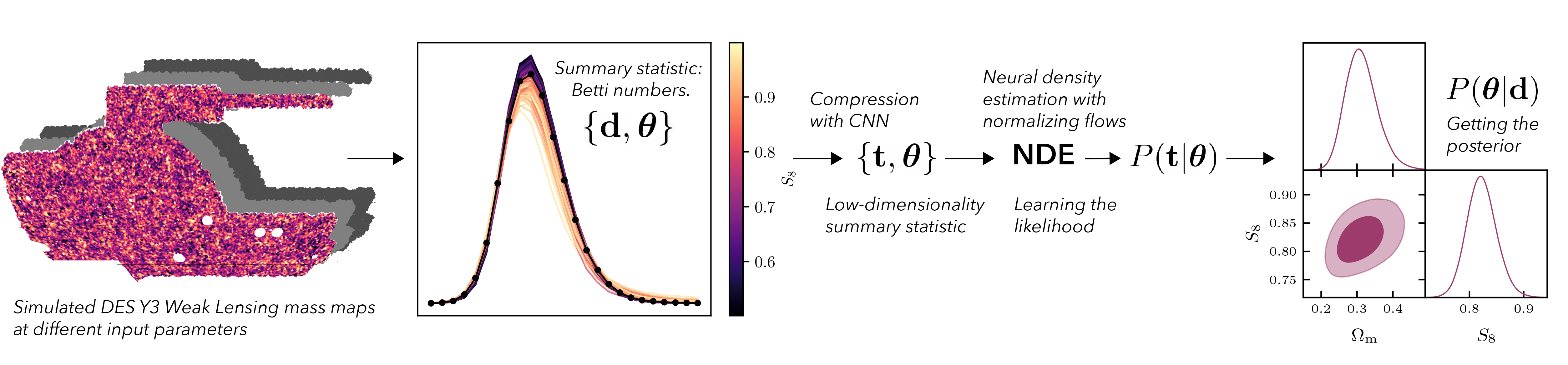}
    \caption{Simulation-based inference (SBI) framework for this work.}
    \label{fig:sbi}
\end{figure*}

\section{Simulation-based inference} \label{sec:sbi}

\subsection{Neural Likelihood Estimation}\label{sec:nle}

Simulation-based inference (SBI, also known as `likelihood-free inference' or `implicit inference') aims to estimate the likelihood, the posterior or the likelihood ratio directly from simulations. We employ the same framework as \citet{Jeffrey2021}, which was extended to DES Y3 in \citet{Jeffrey2025} and \citet{Gatti2024}, where a detailed description of the method is provided. Here we provide a summary.

We use the \textit{neural likelihood estimation} technique, where the conditional probability density function $p(t|\theta)$, commonly referred to as the \textit{likelihood}, is learned from mock data realizations generated through simulation. The process begins by generating samples of $t_i$ according to
\begin{equation} 
t_i \sim p(t | \theta_i) \ . 
\end{equation} 
\noindent Here, $\theta_i$ are the input parameters to the simulation indexed by $i$, generated from a proposal distribution  $\theta_i\sim p_{D}(\theta)$, and $t$ represents the data for which we seek to estimate the likelihood. In our case to generate each $t_i$ we do the following: (1) we construct convergence maps following the methodologies outlined in Sec~\ref{sec:theory} and Sec~\ref{sec:data}; (2) we measure a summary statistic datavector $d$ using persistent homology (see Section ~\ref{sec:persistent-homology}) and (3)
we apply dimensionality reduction to compress datavector  $d$ 
 into a lower-dimensional representation $t$ suitable for likelihood estimation (see Sec.~\ref{sec:compression}). Figure~\ref{fig:sbi} provides a visual representation of this workflow.
 
From the pairs of compressed statistics and input parameters $\{t_i, \theta_i\}$, we learn a density $q$ that approximates $p$, such that $p(t | \theta) \approx q(t | \theta)$. For this purpose, we use the pyDELFI package\footnote{\url{https://github.com/justinalsing/pydelfi}}~\citep{delfi2} to estimate $p(t | \theta)$ with an ensemble of neural density estimators (NDEs). NDEs use neural networks to parameterize densities, including conditional probability densities. An NDE estimates $q(t | \theta, {\varphi})$ by varying the neural network parameters ${\varphi}$ (e.g., weights and biases) to minimize the loss function 
\begin{equation} 
\label{eq:delfi_loss} 
U({\varphi}) = - \sum_{i=1}^N \log q ({t}_i | {\theta}_i ; {\varphi}) \ \ 
\end{equation} 
\noindent over $N$ mock data points $t_i$. This loss function minimizes the expectation of the Kullback-Leibler divergence~\citep{kullback1951} between the true conditional density $p(t|\theta)$ and our approximation $q(t|\theta;\varphi)$ for the range of values of $\theta_i$. \noindent The KL divergence for a given $\theta$ is defined as:
\begin{equation}
D_{\text{KL}}(p(t|\theta) || q(t|\theta;\varphi)) = \int p(t|\theta) \log\left(\frac{p(t|\theta)}{q(t|\theta;\varphi)}\right) dt, 
\end{equation}
\noindent which can be expanded as:
\begin{equation}
D_{\text{KL}} = \int p(t|\theta) \log p(t|\theta) \, dt - \int p(t|\theta) \log q(t|\theta;\varphi) \, dt.
\end{equation}
\noindent When minimizing with respect to $\varphi$, only the second term matters since the first term is independent of $\varphi$. Since each $t_i$ is drawn from $p(t|\theta_i)$, the  loss function in Eq.~(\ref{eq:delfi_loss}) approximates this second term (up to a constant factor $1/N$ that can be omitted during optimization). \noindent This enables our neural density estimator to learn the conditional probability density function $p(t|\theta)$ across parameter space.

For NDEs we use Conditional Masked Autoregressive Flows (MAF; \citealt{MAF}). A MAF, which is a type of Normalizing Flow, uses a series of bijective transformations from simple known densities (e.g., standard Gaussian) to the target density \citep{normalizing_flows,kingma2016improved,MAF2}. Specifically, we use an ensemble of two MAFs, each with two hidden layers of width 50 using tanh activation functions.  The first MAF employs two transformations (Masked Autoencoder for Distribution Estimation, MADE) while the second uses three MADEs. In \citet{Jeffrey2025}  Gaussian Mixture Density Networks (MDN;~\citealt{mdn}) are used to validate the MAFs. 

These neural network architectures have an inductive bias toward learning smooth, continuous probability densities. This architectural constraint makes it computationally difficult for the network to approximate functions like Dirac delta distributions concentrated at specific training points $\theta_i$. Instead, the network naturally learns smooth interpolations between training locations, preventing overfitting to the discrete set of simulation parameters and ensuring generalization across the continuous parameter space.

Once the conditional probability density function $p(t|\theta)$ is estimated it is then evaluated for the observed data $t_O$, relating the posterior probability density of the parameters to the likelihood via Bayes' theorem:
\begin{equation}
p( \theta | t_O ) = \frac{p(t_O | \theta) \ p(\theta)}{p(t_O)} \ \ .
\end{equation}

An important advantage of neural likelihood estimation (as opposed to learning the posterior directly) is that the learned function $p(t|\theta)$ is independent of the proposal distribution $p_D(\theta)$ used to generate the training samples $t_i$. While regions with sparser sampling may have higher uncertainty in the density estimate, the method does not introduce systematic bias based on the training distribution. This flexibility allows the analysis prior $p(\theta)$ used for inference to be different from the proposal distribution $p_D(\theta)$.

\section{Weak gravitational lensing: Mass maps}
\label{sec:theory}

In this work we use convergence (mass) gravitational lensing maps as the  main input data to the analysis. Here we describe how the convergence field is related to the matter density field, and how such maps are obtained from galaxy shape measurements. 

The gravitational potential $\Phi$ and the matter overdensity field $\delta \equiv \delta\rho / \bar{\rho}$ are related by the Poisson equation
\begin{equation}
\label{eq:poisson}
\nabla^2_x \Phi(\boldsymbol{x}, t) = \frac{4 \pi G}{c^2} \bar{\rho}_m (t) a^2(t) \delta(\boldsymbol{x}, t) \ ,
\end{equation}
where $\boldsymbol{x}$ is a comoving spatial coordinate, $t$ is time and $a \equiv 1/(1+z)$ is the scale factor.

Weak gravitational lensing produces small distortions on the image of distant galaxies by mass located between the galaxies and an observer -- see \cite{PratBacon2025} for a review. The effect of weak lensing can be encapsulated in the \textit{lensing potential}, which is an integration of the gravitational potential projected along the line of sight:
\begin{equation}
\label{eq:born}
\phi (\boldsymbol{\theta}, D_S) = \frac{2}{c^2 D_S} \int_0^{D_S} d D_L \frac{D_{LS}}{D_L}\Phi (x^i= D_L \theta^i, D_L) 
\end{equation}
where $D_L$ is the angular distance to the lens, $D_S$ to the source and $D_{LS}$ between the lens and the source, for a single source galaxy.  In practice, we integrate this potential over many source galaxies, using the normalized redshift distribution $n(z)$ of the source galaxy population.

The convergence $\kappa = \kappa_E + i\kappa_B$ (of spin-weight 0 i.e. a scalar) and the shear $\gamma = \gamma_1 + i\gamma_2$ (of spin-weight 2) can be expressed as a function of the lensing potential as
\begin{equation}
\hat{\kappa}_{\ell m}  = - \frac{1}{2} \ell (\ell+1) \hat{\phi}_{\ell m}, \quad \hat{\gamma}_{\ell m} = \frac{1}{2}\sqrt{(\ell-1)\ell(\ell+1)(\ell+2)}\hat{\phi}_{\ell m},
\end{equation}
in the space of spherical harmonics. Thus, in spherical geometry the shear and convergence are related by:
\begin{equation}
\label{eq:mass_map_operator}
\hat{\gamma}_{\ell m} = -\sqrt{\frac{(\ell-1)(\ell+2)}{\ell(\ell+1)}} \hat{\kappa}_{\ell m}.
\end{equation}

The shear ($\gamma$) characterizes distortions in galaxy shapes, while the convergence ($\kappa$) quantifies changes in their apparent sizes. The convergence field is often referred to as a \textit{mass map} since it directly traces the projected matter density along the line of sight. Note that in the decomposition $\kappa = \kappa_E + i\kappa_B$, the $\kappa_B$ component (the ``B-mode'') is expected to be zero in the absence of systematic effects.

In the weak lensing regime, both $\kappa$ and $\gamma$ are much smaller than unity and unfeasible to measure individually for any single galaxy due to the intrinsic variety in galaxy shapes and orientations. However, the lensing signal emerges statistically when averaging shape measurements across many galaxies, as the random intrinsic components average to zero while the coherent lensing effect accumulates. From these averaged shear estimates, we aim to derive the convergence field, which is advantageous both for being a scalar quantity and for providing a more direct probe of mass distribution.

To \textit{reconstruct} the convergence field from shear measurements, we invert the relation presented in Eq.~(\ref{eq:mass_map_operator}). For this work, we implement the spherical reconstruction algorithm developed by \citet{KaiserSquires}, as the flat-sky approximation proves inadequate for the large survey area of DES Y3. The Kaiser-Squires reconstruction requires dividing by $\ell(\ell+1)$ factors in harmonic space, which increasingly amplifies noise at smaller scales (higher $\ell$). Without regularization, this amplification causes the noise to diverge. We therefore apply Gaussian smoothing to suppress the small-scale noise. The specific smoothing scales used in our analysis are detailed in Sec.~\ref{sec:data} and Table~\ref{table:smoothing_scales}. For detailed descriptions of the spherical reconstruction implementation, we direct readers to \citet{Chang2017} and \citet{Jeffrey2025}.

\section{Data and simulations} \label{sec:data}

Our data is from the Dark Energy Survey \citep[DES,][]{DES2016}. DES is an optical near-infrared survey that covers approximately 5000 $\mathrm{deg}^2$ of the southern Galactic sky in five different filters ($grizY$), collecting data from hundreds of millions of distant galaxies up to a redshift of $\sim$1.4. The survey utilizes the Dark Energy Camera \citep{Flaugher2015} on the 4-meter Blanco Telescope at the Cerro Tololo Inter-American Observatory. In this work we use data from the first three years (Y3) of DES observations.

\subsection{Data: DES Y3 Convergence mass maps}

As main input, we use the spherical DES Y3 weak lensing \textit{convergence} map described in detail in \citet*{Jeffrey2021}. In particular, we use the \citet{KaiserSquires} reconstruction since it is a much faster algorithm than the other methods described in \citet*{Jeffrey2021} and it is thus more suitable to run on many mock data for the SBI approach. While this reconstruction method can introduce biases due to survey mask effects, our simulation-based inference methodology naturally accounts for these systematic effects, as we forward-model the entire analysis pipeline—including mask application and reconstruction—identically for both simulations and observed data. 

Such maps are constructed starting from a shape or \textit{shear} catalog, as summarized in Sec.~\ref{sec:theory}. The DES Y3 shape catalog  \citep*{Gatti2021} contains over 100 million galaxies with calibrated galaxy shapes using the \textsc{Metacalibration} algorithm \citep*{Sheldon2017}. It has a weighted galaxy density of $n_{\rm eff}=5.59$~galaxies~arcmin$^{-2}$, over an effective area of 4139  deg$^2$. Further calibration and validation of the shape measurements is performed using image simulations in \citet*{MacCrann2022}. The DES Y3 shape catalog is split into four tomographic redshift bins of similar number density, and each one comes with a  redshift distribution estimated and calibrated based on the framework described in \citet*{Myles2021}. This catalog has been used in cosmological analyses based on two-point correlation functions, both in real space~\citep{Amon2022}, \citet*{Secco2022} and in harmonic space~\citep{Doux2022}.

\subsection{Simulations for SBI: Gower Street suite} \label{sec:sims}

\begin{table}
\caption{Summary of the Gower Street simulation suite. From 791 full-sky shear ($\gamma$) simulations with varying cosmological parameters ($\Omega_i$), we extract four DES footprints per simulation and generate multiple ellipticity ($e$) maps with different nuisance parameters ($n_i$) and shape noise realizations ($e_\mathrm{SN}$). These are then transformed into convergence maps via KS reconstruction.}
    \centering
    \begin{tabular}{cc c}
        \toprule
         & $N$ & Input varied \\
        \midrule
        Full-sky $\gamma-$maps &  791 & $\Omega_i$  \\
        Independent DES $e-$maps & 791 $\times$ 4 = 3164 & $n_i$ \\
        Quasi-independent DES $e-$maps & 3164 $\times$ 4 = 12656 &  $n_i$, $e_\mathrm{SN}$ \\
        \bottomrule
    \end{tabular}
    
    \label{tab:sims_summary}
\end{table}

\begin{table}
\caption{Distribution of parameters used in the Gower Street simulation suite and their corresponding analysis prior when sampling the likelihood. $\mathcal{N}(\mu,\sigma)$ denotes a normal distribution with the indicated mean and standard deviation and $\mathcal{U}[a, b]$ denotes a uniform distribution with the indicated limits.} 
\centering
\begin{tabular}{lp{4cm}p{2cm}}
\toprule
\textbf{Parameter} & \textbf{Distribution} & \textbf{Analysis prior} \\
\midrule
\multicolumn{3}{l}{\textit{Cosmological parameters}} \\
\midrule
$\Omega_{\rm m}$ & Active learning $\mathcal{U}(0.15,0.49)$ & $\mathcal{U}(0.15,0.49)$ \\
$S_8$ & Active learning  $\mathcal{U}(0.5,1.0)$ & $\mathcal{U}(0.5,1.0)$ \\
$w$ & $\mathcal{N}(-1,\frac{1}{3})$ in $[-1,-\frac{1}{3})^{\dagger}$ & $\mathcal{U}(-1,-\frac{1}{3})$ \\
$n_s$ & $\mathcal{N}(0.9649, 0.0063)$ &  $-$ \\
$h$ & $\mathcal{N}(0.7022, 0.0245)$ &  $-$ \\
$\Omega_{\rm b}h^2$ & $\mathcal{N}(0.02237, 0.00015)$ &  $-$ \\
$\log(m_{\nu})$ & $\mathcal{U}[\log(0.06), \log(0.14)]$ &  $-$ \\
\midrule
\multicolumn{3}{l}{\textit{Nuisance parameters}} \\
\midrule
$A_\mathrm{IA}$ & $\mathcal{U}[-3, 3]$ & $\mathcal{U}[-3, 3]$ \\
$\eta_\mathrm{IA}$ & $\mathcal{U}[-5, 5]$ & $-$  \\
$m_{1}$ & $\mathcal{N}(-0.0063, 0.0091)$ &  $-$ \\
$m_{2}$ & $\mathcal{N}(-0.0198, 0.0078)$ &  $-$ \\
$m_{3}$ & $\mathcal{N}(-0.0241, 0.0076)$ &  $-$ \\
$m_{4}$ & $\mathcal{N}(-0.0369, 0.0076)$ &  $-$ \\
$\bar{n}_i(z)$ & $p_{\textsc{HyperRank}}(\bar{n}_i(z) | x_{\rm phot})$ & $-$ \\
\bottomrule
\multicolumn{3}{l}{$^{\dagger}$For 64 simulations values outside the range were not discarded.} \\
\end{tabular}
\label{tab:sims_pars}
\end{table}

We base our inference methodology described in Sec.~\ref{sec:sbi} on the  Gower Street (Gower St) simulation suite, which has been built to specifically mimic the properties of the DES Y3 convergence maps. Here we summarize its main characteristics and refer the reader to \citet{Gatti2024} and \citet{Jeffrey2025} for further details and validation of the suite. 

The Gower St simulation set consists of 791 gravity-only full-sky $N$-body simulations, produced using the \textsc{PKDGRAV3} code \citep{potter2017pkdgrav3}. For each full sky shear map we cut four DES footprints, for which we draw four shape noise realizations, as detailed in Table~\ref{tab:sims_summary}. The simulations span a seven-dimensional parameter space in $\nu w$CDM, with the cosmological parameters listed in Table~\ref{tab:sims_pars}. $\Omega_{\rm m}$ and $\sigma_{\rm 8}$ have been sampled with a mixed active-learning strategy. In particular they were at first distributed according to the existing DES analysis constraints, and then after an initial simple blind power spectrum analysis, new simulations were run with $\sigma_8$ and $\Omega_{\rm m}$ values (known only to the computer) in regions of parameter space with poor accuracy of the likelihood estimates (see \citealt{Jeffrey2025}). The other cosmological parameters were chosen to be distributed as follows:
\begin{itemize}
\item{$n_s \sim \mathcal{N}(0.9649, 0.0063)$; from Planck \citep{aghanim2020planck} with the standard deviation boosted by a factor of 1.5.}
\item{$h \sim \mathcal{N}(0.7022, 0.0245)$; consistent with both SH0ES~\citep{Riess_2022} and Planck \citep{aghanim2020planck}.}
\item{$\Omega_{\rm b}h^2 \sim N(0.02237, 0.00015)$; from Planck \citep{aghanim2020planck}.}
\item{$w \sim \mathcal{N}(-1, 1/3)$, but with values less than $-1$ or greater than $-1/3$ then discarded. For a few (64) simulations, part of the ``science verification'' runs, this discarding was not done. }
\item{$m_{\nu}$: fixed at $0.06$ for 192 simulations and with $\log(m_{\nu}) \sim \mathcal{U}[\log(0.06), \log(0.14)]$ thereafter.}
\end{itemize}
In the above, $\mathcal{N}(\mu,\sigma)$ denotes a normal distribution with the indicated mean and standard deviation and $\mathcal{U}[a, b]$ denotes a uniform distribution with the indicated limits.

Additionally, the Gower Street simulation suite incorporates several classes of nuisance parameters to account for systematic uncertainties. For intrinsic alignments, we vary both the amplitude ($A_{\rm IA}$) and redshift scaling ($\eta_{\rm IA}$) parameters under the non-linear alignment model, while also accounting for the interplay between source clustering and intrinsic alignments (\citealt{Gatti2024} provides detailed definitions and treatment). We include variations in the multiplicative shear bias parameters $m_i$ (defined in \citealt{MacCrann2022}). Additionally, we incorporate redshift distribution uncertainties using the samples generated by \citet*{Myles2021} that marginalize over the full shape of the redshift distributions. Precisely, an advantage of simulation-based inference is its ability to easily marginalize over these high-dimensional nuisance parameter spaces by varying them directly in the input simulations.

\subsection{Simulations for some validation tests: CosmoGrid} \label{sec:cosmogrid}

We use a subset of the simulations from the \texttt{CosmoGridV1} suite \citep{cosmogrid1} for additional testing and to assess baryonic contamination. The \texttt{CosmoGridV1} simulations have also been produced using the \textsc{PKDGRAV3} code. From the available \texttt{CosmoGridV1} simulations we chose a set of 200 full-sky simulations at the fiducial cosmology $\sigma_8 = 0.84$, $\Omega_{\rm m}=0.26$, $w=-1$, $H_0=67.36$, $\Omega_{\rm b}=0.0493$, $n_{\rm s}=0.9649$. Each individual simulation has also been post-processed with a baryonification algorithm that mimics the impact of baryons at small scales. 
The algorithm used is the baryonic correction model \citep{Schneider2019,Fluri2022}, which adjusts the particle positions in gravity-only simulations to mimic the impact of various baryonic processes on the density distribution. The cosmology has been chosen to be well-centered within our priors for $\sigma_8$ and $\Omega_{\rm m}$.

\begin{table}
\caption{Parameter values we use for processing the maps for capturing different scales. The pixel size and Gaussian smoothing scale ($\theta_s$, corresponding to the full width half maximum, FWHM) are in arcminutes. The $N_\text{side}$ value corresponds to the target \textsc{HEALPix}  resolution of the map, after downgrading following the procedure described in Sec.~\ref{sec:processing_maps}.}
\centering
\begin{tabular}{|l|c|c|c|c|c|c|c|c|}
\hline
\textbf{$N_\mathrm{side}$} & 512 & 512 & 512 & 512 & 256 & 128 & 128 & 64 \\
Pixel size  & 6.9 & 6.9 & 6.9 & 6.9 & 13.7 & 27.5 & 27.5 & 55.0 \\
$\theta_s$ & 8.2 & 13.1 & 21 & 33.6 & 54 & 86 & 138 & 221 \\
\hline
\end{tabular}
\label{table:smoothing_scales}
\end{table}

\section{Summary statistic: Persistent homology}  \label{sec:persistent-homology}

\begin{figure}
   \centering
    \includegraphics[width=0.5\textwidth]{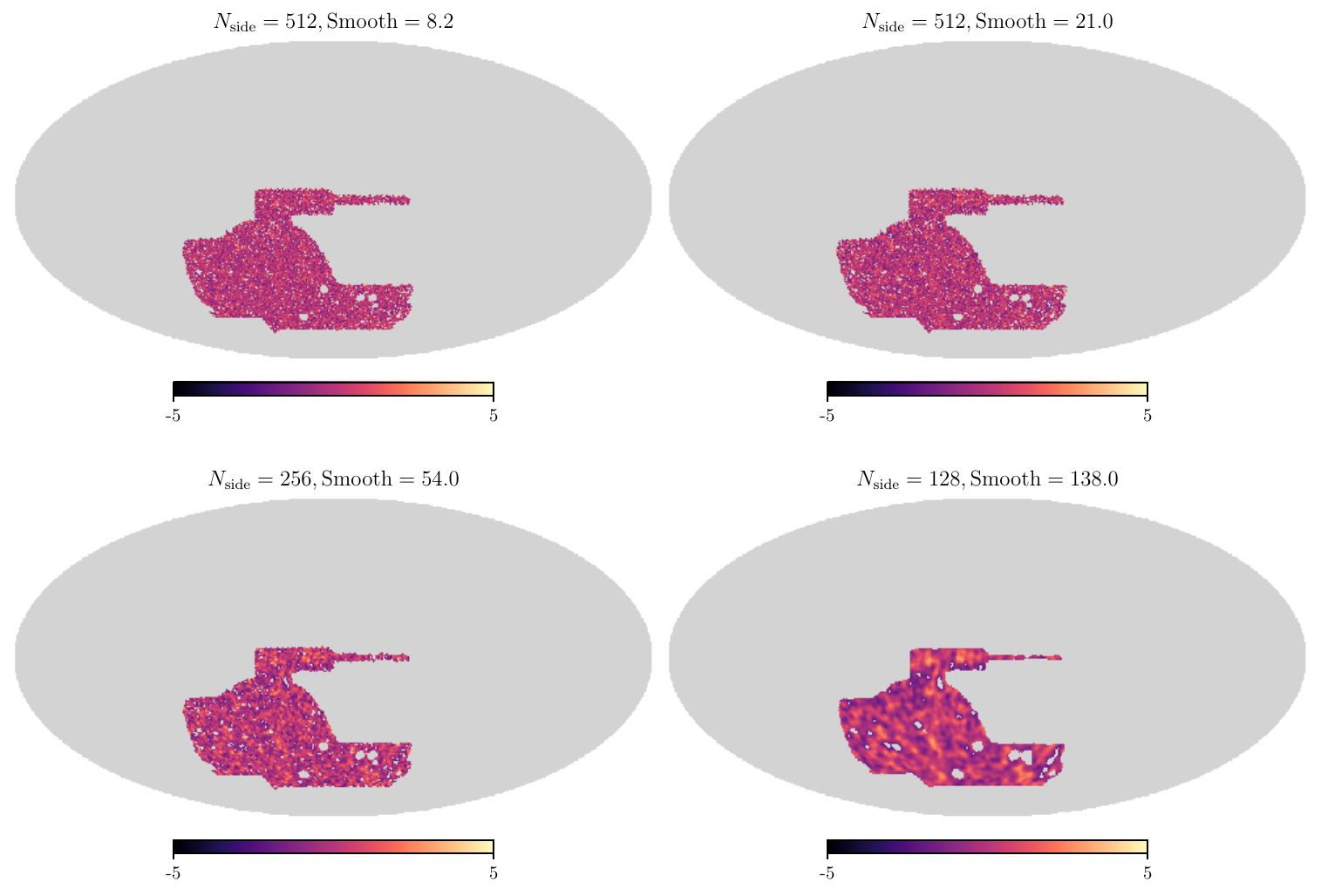}
    \caption{Processed DES Y3 convergence maps following the steps (i) to (iii) from Sec.~\ref{sec:processing_maps}:  smoothing and downgrading the map, masking and rescaling it.  This plot is showing the data from the highest redshift bin.}
    \label{fig:processed_maps}
\end{figure}

Homology is a mathematical formalism that categorizes topological spaces by  unambiguously and exactly detecting the independent holes of different dimensions they contain. A $D$-dimensional manifold may contain holes of $0$ up to $D$-dimensions. The $p$-dimensional holes of a manifold are  associated with its $p$-th homology group, and the number of independent $p$-dimensional holes,  captured by the rank of the Homology group, denoted as the Betti number $\beta_p$.  For the spherical manifold $\mathbb{S}^2$, only the $0$- and $1$-dimensional homology groups are relevant. Three-dimensional datasets typically also consider $H_2$ (cavities). For $\mathbb{S}^2$, there exists only one such cavity—the interior of the sphere itself.

However, our data is masked with regions in $\mathbb{S}^2$ where the data is missing or spurious. We therefore apply \textit{relative homology}, computing the topological features of the unmasked regions while treating mask boundaries as topological constraints. This relative homology framework analyzes the pair (unmasked region, mask boundary), properly accounting for how masking affects connectivity and hole structure. It is important to note that while the topology of the mask is a constant for a 
given mask, how its features affect the topology of the non-masked region is dictated by a particular realization of the field, and therefore the same mask may not have the same effect across different realizations of a model. Also, even though $H_2$ is not relevant in standard homology over $\mathbb{S}^2$, this is not the case in relative homology where such features can form. Still, for our current analysis, we focus exclusively on the $H_0$ and $H_1$ homology groups, as $H_2$ features are substantially less numerous and unlikely to provide significant constraining power. Future work could investigate whether including $H_2$ features yields additional cosmological information. Then, in this work we consider the following relative homology groups:
\begin{itemize}
\item \textbf{Connected components} ($H_0$): regions with filled values that are contiguous inside the mask.  In other words, for any two points of a connected component, it is possible to define a continuous path between them.\\
\item \textbf{Holes} ($H_1$): empty regions encircled by connected structures in the convergence field, including voids created by masked areas. 
\end{itemize}

\textit{Persistent homology} extends homology by computing topological features across multiple threshold levels, enabling the identification of birth and death thresholds for each topological structure in the field. This approach associates a persistence measure with each feature---the range of thresholds over which it exists---providing a natural significance metric. While a conventional interpretation suggests more  persistent features are detected over a wide range of map values and are deemed more likely to represent true features of the underlying space rather than artifacts of sampling, noise, or particular choice of parameters, in hierarchical and multiscale fields, the interpretation is more nuanced.

\subsection{Computational pipeline} \label{sec:computation}

In this work we use \textsc{TopoS2} \citep{Pranav2022}, a tool designed specifically for computing persistent homology on spherical geometries. TopoS2 is developed in C++, offering the advantage of rapid computations --- essential for simulation-based inference.  It is built to work with the \textsc{HEALPix} \citep{GORSKI2005} framework, since it was developed to be applied to CMB maps.  See \citet{Pranav2019} and \citet{pranav2025} for extended details on how the computational process  is performed on \textsc{HEALPix} maps and how the triangulation is done. In this section we summarize the process that \textsc{TopoS2} performs to measure persistent homology. 

\subsubsection{Pre-processing the original convergence maps} \label{sec:processing_maps}

\begin{figure}
\centering
\begin{subfigure}[b]{0.5\textwidth}
\centering
   \includegraphics[width=0.85\linewidth]{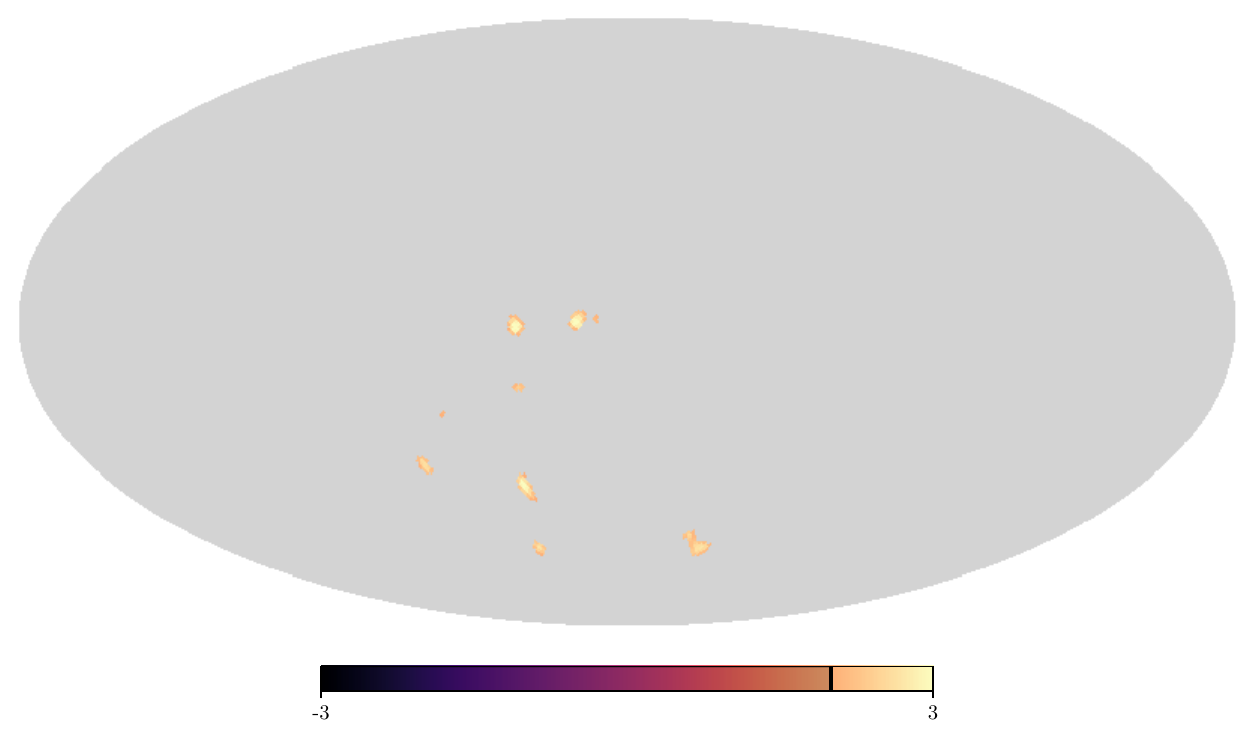}
   \label{fig:Sub3}
\end{subfigure}

\begin{subfigure}[b]{0.5\textwidth}
\centering
   \includegraphics[width=0.85\linewidth]{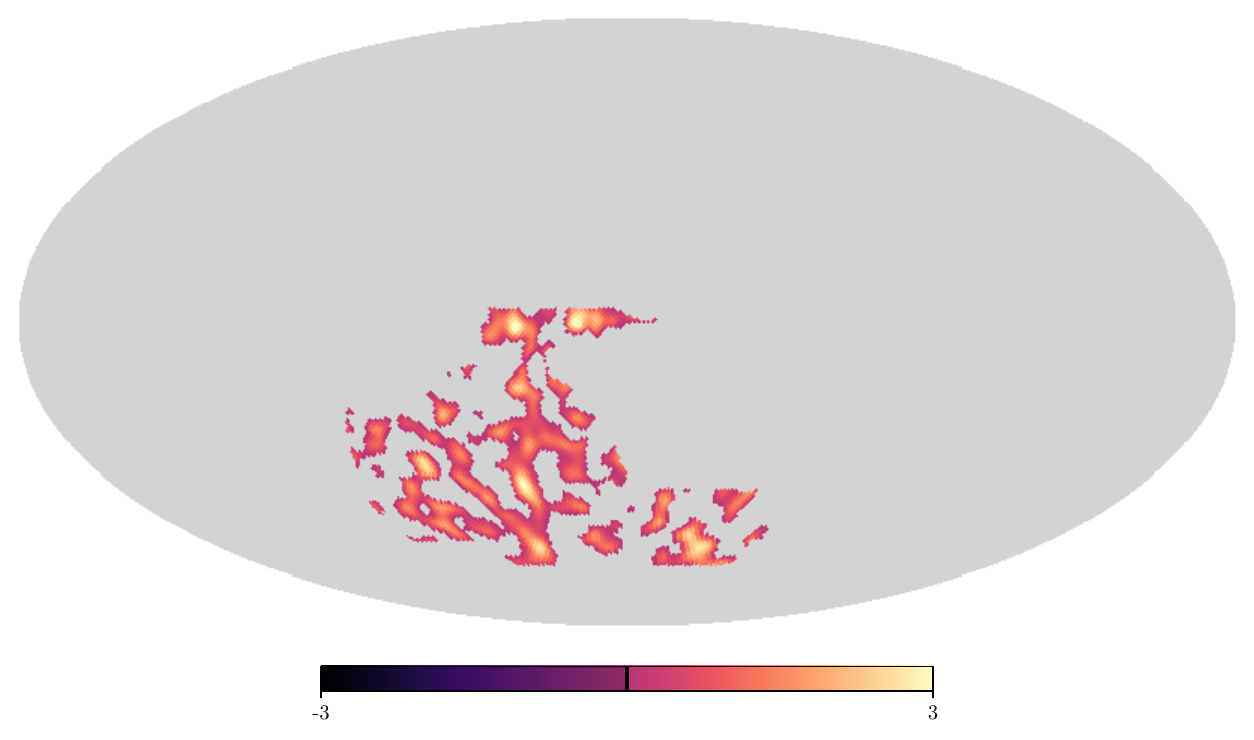}
   \label{fig:Sub2}
\end{subfigure}

\begin{subfigure}[b]{0.5\textwidth}
\centering
   \includegraphics[width=0.85\linewidth]{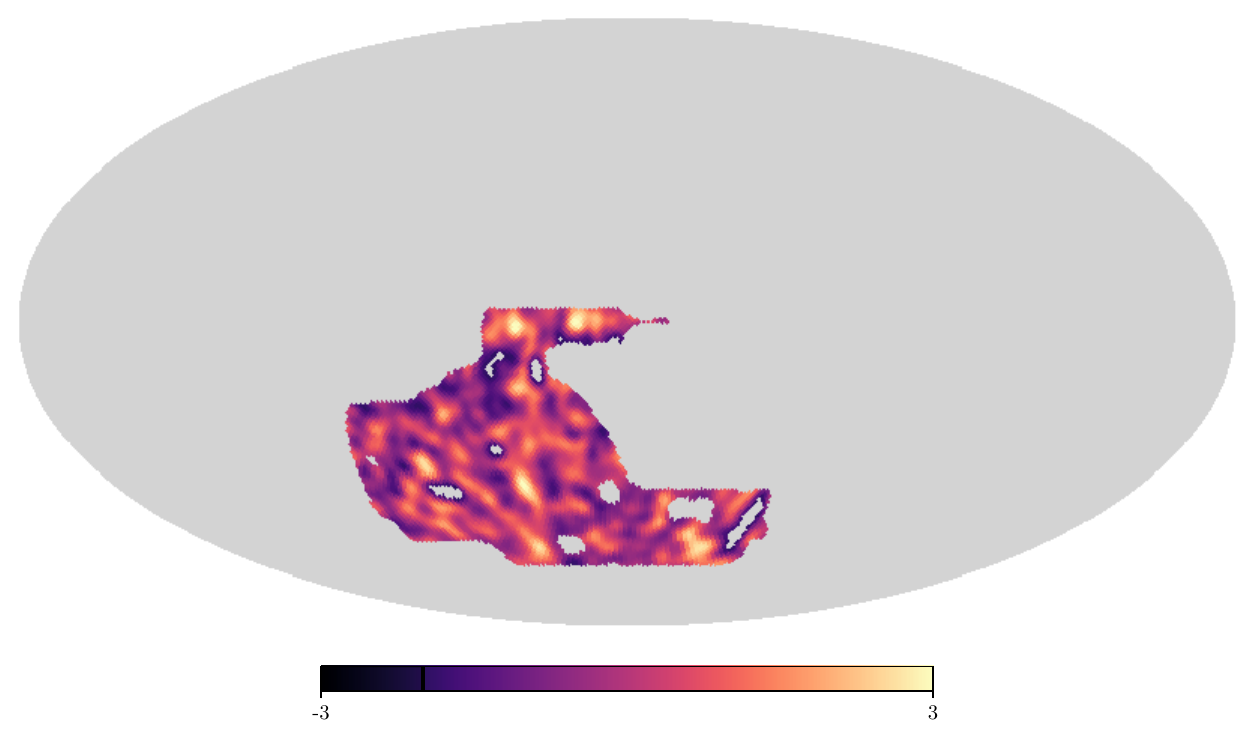}
   \label{fig:Sub1}
\end{subfigure}

\caption{Visualization of a superlevel set. The colorbar represents $\kappa_r$, the rescaled convergence field as defined in Eq.~(\ref{eq:nu}). The top panel shows regions where $\kappa_r>2$, revealing mostly connected components that correspond to  peaks. In the middle panel ($\kappa_r>0$), both connected components and holes are visible. The lower panel ($\kappa_r>-2$) only has holes, corresponding to underdense regions in the density field. This figure depicts the DES Y3 highest redshift bin, smoothed with $\theta_s=221$ arcmin, our largest scale.}
\label{fig:filtration}
\end{figure}

\begin{figure}
   \centering
    \includegraphics[width=0.41\textwidth]{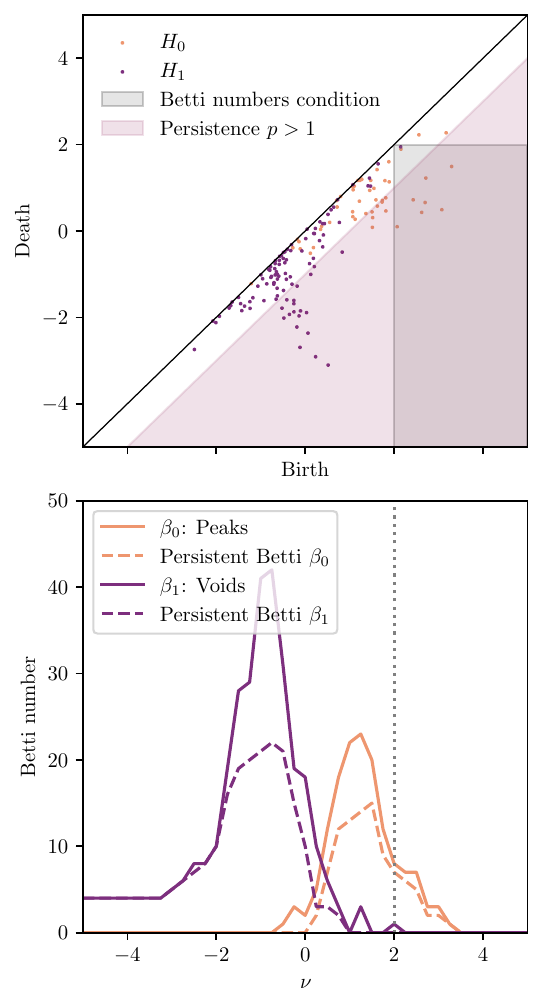}
    \caption{Persistence diagram and (persistent) Betti numbers corresponding to the superlevel set in Fig.~\ref{fig:filtration}. The connected components ($H_0$ homology group) are represented in orange and the holes ($H_1$) in purple. Note birth $>$ death because we construct the superlevel set beginning at the highest threshold and descend due to its computational advantages. Betti numbers count the points in the persistence diagram within a certain region. For $\nu=2$, we represent such region with a gray shaded box, where 8 points can be counted for $H_0$, matching the $\beta_0 (\nu=2)$ value in the lower panel and the number of connected components in the top panel of Fig.~\ref{fig:filtration} minus one (the component that is born at the highest threshold never dies and is mapped to the single connected surface when the excursion set is finished). Persistent Betti numbers apply an additional condition that removes \textit{less persistent} points (those more likely attributable to noise) by requiring that the persistence $p$ (the difference between death and birth) exceeds a threshold. In this work, we use $p>1$, represented by the purple shading in the top panel.}   
    \label{fig:betti_illustrated}
\end{figure}

\begin{figure*}
   \centering
    \includegraphics[width=1\textwidth]{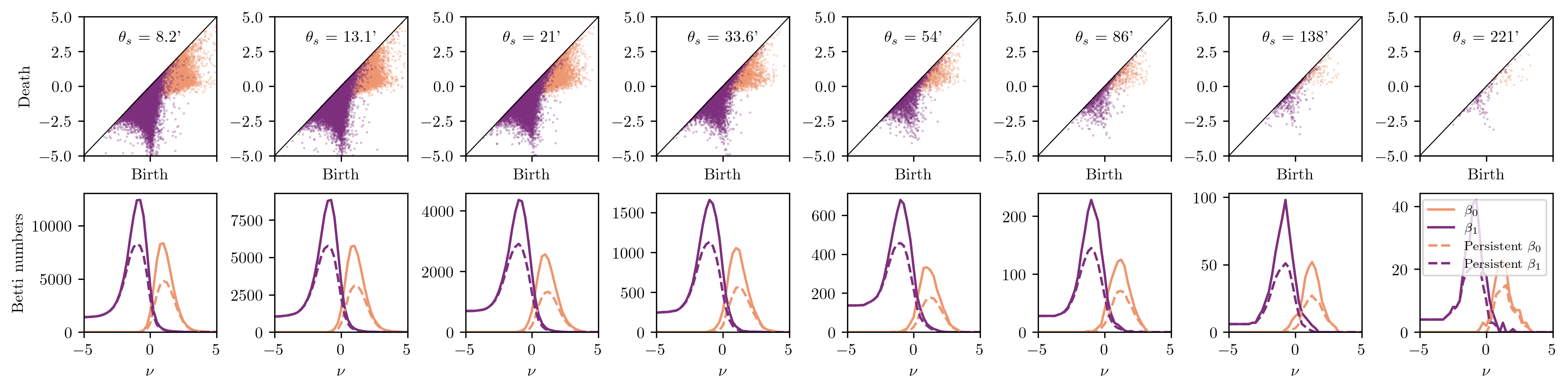}
    \caption{Persistence diagrams and corresponding Betti numbers (solid lines) and persistent Betti numbers (dashed lines) for the DES Y3 data for all smoothing scales, for the highest redshift bin. Our analysis represents the first multi-scale persistent homology analysis applied to cosmological data.}
    \label{fig:data-vector}
\end{figure*}

Before building the persistence diagram, the following transformations are applied to the maps:
\begin{enumerate}
    \item \textbf{Smooth and downgrade}: We extract the spherical harmonic coefficients at the input resolution, convolve with a Gaussian filter of a given FWHM (smoothing scale), and resynthesize the map at the desired resolution from the Gaussian convolved harmonic coefficients.
    Lower resolutions (smaller $n_\text{side}$ values) are significantly less expensive to compute, so we downgrade to the minimum resolution that is suitable for a given smoothing. The computational time ranges from a few seconds for the fastest configuration to approximately 8 minutes per map for the slowest configuration using a single core. We  choose resolutions whose pixel size is at least three times smaller than the smoothing scale we plan to apply (except in the smallest smoothing scales, where going to a higher resolution would make the computations too expensive). We list all the resolution and smoothing scales in Table~\ref{table:smoothing_scales}. We also implement a hard cutoff in harmonic space at $\ell_{\text{max}} = 2 N_\mathrm{side}$. 
    \item \textbf{Mask}: We smooth the mask with the same scale, and rebinnarize it with a chosen threshold. We choose to keep all the pixels whose value is above 0.8. We apply this binary mask to the map. 
    
    \item \textbf{Rescale}: We then rescale the original convergence map $\kappa$ in the following way:
            \begin{equation}\label{eq:nu}
                \kappa_r = \frac{\kappa - \mu_\kappa }{\sigma_\kappa},
            \end{equation}
            where $\kappa_r$ and $\kappa$ represent the values of the field at each pixel, and $\mu_\kappa$ and $\sigma_\kappa$ represent the mean and standard deviation of the convergence field across all pixel values inside the mask. We show the output of such process in Fig.~\ref{fig:processed_maps}, for all smoothing scales used in our analysis.
            While this rescaling preserves the persistence diagram structure for individual maps, it has important implications when analyzing an ensemble. Though $\mu_\kappa$ remains cosmologically uninformative (approximately zero), $\sigma_\kappa$ contains significant cosmological information. By standardizing, we largely separate the second-moment contribution from the persistent homology signal. This approach allows us to either analyze the combined signal (persistent homology plus second moments) or focus primarily on the higher-order information. We explore this distinction further in Sec.~\ref{sec:2ndmoments}.
\end{enumerate}

\subsubsection{Building a persistence diagram} \label{sec:build_diagrams}
Having processed the maps as outlined in the previous section, we are ready to compute the persistent homology statistic. To generate a persistence diagram, the following  steps are followed:

\begin{enumerate}
    \item \textbf{Building a filtration from excursion sets}. An \textit{excursion set} or a \textit{superlevel set} of a scalar field $\kappa_r(\mathbf{x})$ on the sphere $\mathbb{S}^2$ is defined  as 
    \begin{equation}\label{eq:threshold}
        \mathbb{E} (\nu) = \{ \mathbf{x} \in \mathbb{S}^2 \, | \, \kappa_r(\mathbf{x}) \geq \nu \} 
    \end{equation}
    where $\mathbf{x}$ represents each \textsc{HEALPix} pixel. Fig.~\ref{fig:filtration} presents a visualization of the excursion sets of the DES Y3 convergence map for a given redshift bin. We analyze the map across multiple thresholds by building a nested sequence of  excursion sets, which is known as a \textit{filtration}. A filtration may also be built from \textit{sublevel sets} of a function. In our case, superlevel and sublevel filtrations yield equivalent  information, with the homology groups $H_0$ and $H_1$  and birth and death values interchanged. However, the superlevel filtration is more meaningful in cosmological applications as this approach associates topological islands with density peaks, and the topological holes are associated with the regions around the density minima \citep{Feldbrugge2019}.
    \item \textbf{Record features from each homology group}. We track the emergence and disappearance of topological features, specifically \textit{connected components} ($H_0$ homology group) and \textit{holes} ($H_1$ homology group), as a function of threshold ($\nu$). Due to computational advantages, in superlevel sets the filtration is usually started at the highest threshold value, and it gets lower subsequently. In that case, connected components emerge first, corresponding to the highest peaks in the density field. When a feature first appears, we record this threshold as its \textit{birth}. Conversely, when a feature disappears, we record this as its \textit{death}, and a point is placed in the persistent diagram, as shown in the upper panel of Fig~\ref{fig:betti_illustrated}.  When two components merge, the older component (born at an earlier threshold) persists, while the younger one is considered to have died.  
\end{enumerate}

\subsection{Cross-tomographic measurements}

We also include the measurements of cross-correlations between different redshift bins, since these carry a significant amount of information \citep{Martinet2018}. We combine the convergence maps for all possible two-, three-, and four-map configurations, then apply the methodology detailed in Sections \ref{sec:processing_maps} and \ref{sec:build_diagrams}. To combine the  maps we simply add\footnote{While alternative combination methods exist, current literature shows no clear consensus regarding their relative advantages.} the convergence from the individual redshift bins, as done in e.g. \citet{Cheng2025}. Adding the cross-correlations significantly improves the constraints on both $\Omega_{\rm m}$ and $S_8$.

\subsection{Betti numbers}

A persistence diagram can be compressed to the so-called Betti numbers\footnote{While this compression is not lossless, it transforms the two-dimensional diagram into a 1D vector that proved significantly more amenable to neural network processing, which we discuss in Sec.~\ref{sec:compression}. Despite exploring various architectures, direct compression of full persistence diagrams remained challenging—the resulting models consistently underperformed compared to those using Betti numbers. 2D diagrams require substantially more memory and computational resources than Betti numbers, making the latter a pragmatic choice that balances information retention with implementation feasibility. We leave the development of architectures that effectively compress 2D diagrams for future work.}. The Betti numbers count the number of  features of the excursion set. In our case, $\beta_0$ counts the number of connected components and $\beta_1$  the number of holes.  The Betti numbers $\beta_i(\nu)$ can therefore be obtained by summing over all events with birth, death $(a, b)$ with $a \geq \nu$ and $b < \nu$, as illustrated in Fig.~\ref{fig:betti_illustrated}.

We calculate Betti numbers relative to the mask at intervals of 0.25 in $\nu$ (ranging from -5 to 5), generating 41 data points for each tomographic bin combination and homology group. Our complete datavector encompasses 15 redshift bin combinations (4 auto-correlations from the original bins, 6 pair-wise correlations, 4 three-way combinations, and the full combination of all bins) $\times$ 41 threshold points $\times$  2 homology groups $\times$  8 smoothing scales, yielding a datavector with 9,840 data points. In Fig.~\ref{fig:data-vector} we show the  measurements for a given DES Y3 redshift bin.

\subsection{Persistent Betti numbers}

Persistent Betti numbers are defined similarly to Betti numbers, with an additional condition. Specifically, we require that the persistence \( p \), defined as the difference between the birth and death values $ p \equiv b - d $, exceeds a given threshold. For this work we use $p=1$\footnote{This choice is to some extent arbitrary, however we observe in Fig.~\ref{fig:2d_histograms} that the region of $1 \lesssim p \lesssim 3$ exhibits the strongest signal-to-noise ratio for our case.}. Figure~\ref{fig:betti_illustrated} illustrates the region over which one should sum to obtain the corresponding persistent Betti number at a specific threshold \( \nu \), depicted as the purple shaded area. In Fig.~\ref{fig:data-vector} we show how they compare with the Betti numbers for a given DES Y3 redshift bin.

Contrary to our initial expectation that persistent Betti numbers would yield tighter constraints by filtering noise, both statistics perform similarly in our analysis (see Fig.~\ref{fig:betti_vs_moments_comparison}). Figure~\ref{fig:2d_histograms} helps explain this outcome --- while the signal-to-noise ratio peaks at persistence values around $p\sim2$, substantial cosmological information remains in features with persistence below our threshold. This suggests our $p=1$ cutoff simultaneously removes both noise and informative topological features. 

The comparable performance indicates that our compression process (detailed in Sec.~\ref{sec:compression}) successfully identifies cosmologically relevant information regardless of whether low-persistence features are included. To characterize the intrinsic performance of both statistics under optimal conditions, we conducted this comparison using a nearly noiseless input created by averaging 800 \texttt{CosmoGrid} simulations with identical cosmological parameters.

\subsection{Including second moments} \label{sec:2ndmoments}

\begin{figure*}
    \centering
    \begin{minipage}{0.38\textwidth}
        \centering
        \includegraphics[width=\textwidth]{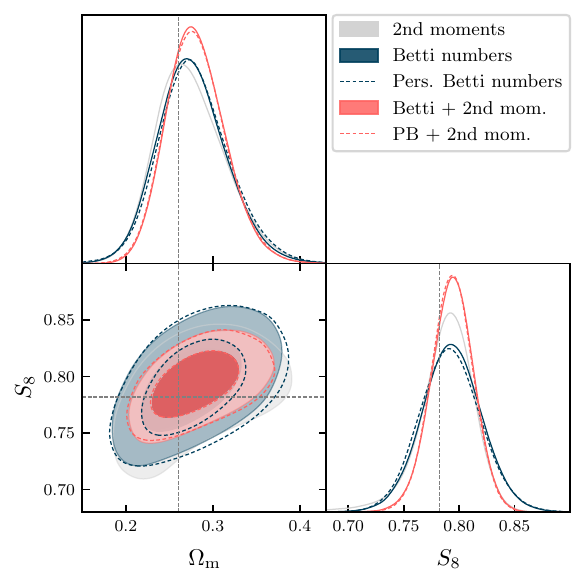}
    \end{minipage}
    \hfill
    \begin{minipage}{0.58\textwidth}
        \centering
        \small
    \begin{tabular}{lcccc}
    \toprule
    \textbf{\texttt{CosmoGrid}} & $\sigma(S_8)$ & $\sigma(\Omega_\mathrm{m})$ & $\sigma(A_\mathrm{IA})$ & \textbf{FoM ($\Omega_\mathrm{m}$, $S_8$)}  \\
    & $[\times 100]$ & $[\times 100]$ & $[\times 10]$ & \\
    \midrule
    $\beta_0$ (Clusters) & 2.9 & 4.4 & 4.1 & 836 \\
    $\beta_1$ (Voids) & 3.0 & 4.6 & 4.3 & 718 \\
    \midrule
    2nd moments & 2.3 & 4.0 & 4.0 & 1053 \\
    \midrule
    Betti numbers & 2.7 & 3.9 & 3.6 & 1060 \\
    Betti + 2nd moments & 1.9 & 3.3 & 3.1 & 1732 \\
    & \textit{(+42\%)} & \textit{(+18\%)} & \textit{(+16\%)} & \textit{(+63\%)} \\
    \midrule
    Persistent Betti numbers  & 2.7 & 4.0 & 4.0 & 984 \\
    Persistent Betti + 2nd moments & 1.9 & 3.4 & 3.2 & 1707 \\
    & \textit{(+42\%)} & \textit{(+18\%)} & \textit{(+25\%)} & \textit{(+73\%)} \\
    \bottomrule
    \end{tabular}
    \end{minipage}
    \caption{Comparison between constraints using different summary statistics for the \texttt{CosmoGrid} simulations. \textbf{Left:} Constraints in the $\Omega_\mathrm{m}$-$S_8$ plane using 2nd moments only, Betti numbers only, and their combination. The solid lines represent standard Betti numbers while dashed lines show persistent Betti numbers. \textbf{Right:} Quantitative comparison showing parameter uncertainties and Figure of Merit improvements. The Figure of Merit is given by FoM = $(\mathrm{det}[\mathrm{Cov}])^{-1/2}$ for the posterior covariance and is a measure of inverse volume (i.e. tightness) of the posterior probability. The percentage improvements in parentheses show the improvement when 2nd moments are added. }
    \label{fig:betti_vs_moments_comparison}
\end{figure*}

We rescale each convergence map by its standard deviation to normalize the scaling of the Betti numbers, as outlined in Section \ref{sec:processing_maps}. This rescaling separates the cosmological information contained in the overall amplitude (i.e. the \textit{second moments}, which contain similar information to the power spectrum) from the topological features, enforcing that the statistic primarily captures higher-order (beyond ``Gaussian'').  We have the option to reintroduce this Gaussian (2-point) information via the concatenation of the second moments to the Betti numbers data vector.

For the second moments the length of the datavector is 120, corresponding to a single number -- the standard deviation --  for each redshift bin ($\times$15) and each smoothing scale ($\times$8). The inclusion of second moments in our analysis leads to notable improvements in cosmological parameter constraints. Specifically, the addition of 2nd moment statistics tightens constraints on both $\Omega_\mathrm{m}$ and $S_8$ parameters, with the Figure of Merit improving from 1053 to 1732  for  standard Betti numbers and from 984 to 1707  for persistent Betti numbers (about a 70\% increase), as seen in Fig.~\ref{fig:betti_vs_moments_comparison} for the \texttt{CosmoGrid} simulations.

\section{Compression} \label{sec:compression}

The complete data vector from our persistent homology statistics has high dimensionality (9,840 data points for the Betti numbers), making direct likelihood estimation computationally intractable within the SBI framework described in Sec.~\ref{sec:sbi}. We implement neural compression to transform these statistics into low-dimensional representations that we can use for parameter inference. 

It is important to point out that poor compression (losing information) will not lead to biased inference. Consistent application of the same compression to both simulated and observed data means a less-informative summary statistic  results in inflated posterior distributions on $\theta$. However, although poor compression will not lead to incorrect inference, finding a compression scheme that maximizes information about $\theta$ is ideal. 

Below we describe the details of such compression for each input statistic.

\begin{figure}
    \centering
    \includegraphics[width=0.45\textwidth]{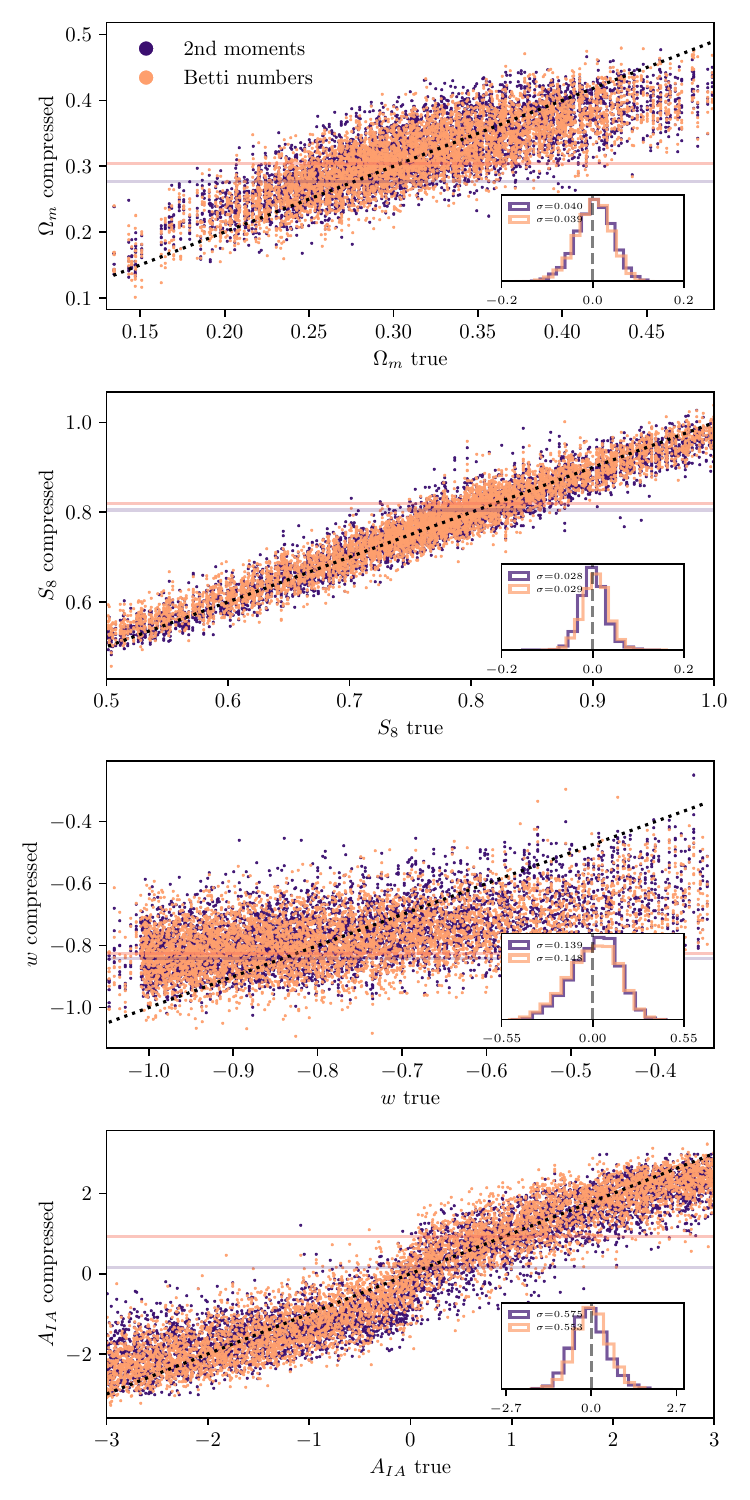}
    \caption{Comparison between true and compressed cosmological parameters. Each panel shows the scatter plot between the true parameters  on the $x$-axis and their compressed values on the $y$-axis.  The dashed line indicates perfect recovery. The horizontal lines represent the compressed values for the data realization, for which the truth is unknown. Inset histograms show the distribution of residuals (compressed - true) for each parameter. }
    \label{fig:parameter_compression}
\end{figure}

\subsection{Network architecture for the Betti numbers}

We use separate neural networks for each target parameter ($\Omega_\mathrm{m}$, $S_8$, $w$, and $A_\mathrm{IA}$). We train with mean-squared error (MSE) loss to approximate posterior means. These networks convert our topological statistics into parameter point estimates that serve as compact summary statistics for subsequent neural likelihood estimation. In Fig.~\ref{fig:parameter_compression} we show the result of such compression.

For the compression of the Betti numbers, we employ a specialized convolutional neural network (CNN) architecture designed to capture the correlations between smoothing scales and redshift bins. The architecture processes the normalized Betti numbers with dimensions (40, 240), where 240 represents the 15 tomographic bin combinations × the 2 homology groups × the 8 smoothing scales, and 40 represents the length of each Betti curve\footnote{We drop one data-point from each Betti curve to achieve an even-dimensional input for the CNN architecture. Since this data-point typically approaches zero in most cases, its omission has negligible impact on the information content of our analysis.}. After experimenting with various convolutional approaches, we selected \texttt{SeparableConv2D} layers as the primary building blocks of our network. This choice is particularly effective for our data structure, as \texttt{SeparableConv2D} processes each input channel independently before mixing information across channels. In our case,  we reshape the input to (40, 8, 30), where the 30 channels correspond to the various tomographic bins and homology groups, allowing the network to initially learn features specific to each redshift bin configuration and topological structure (clusters or voids) separately. This channel-wise processing significantly reduces the number of parameters while preserving the network's ability to capture meaningful cosmological information across different scales. The subsequent point-wise convolution then efficiently combines these channel-specific features. We provide complete details on the implementation in Appendix~\ref{app:architecture}.

Before training the compression network, we apply standardization to improve training stability and convergence. This preprocessing is applied consistently to both training and validation data, as well as to the observed data during inference.

To improve the training, we also implement:

\begin{itemize}
    \item An adaptive learning rate schedule starting at $10^{-3}$ with reduction when the validation loss plateaus, as proposed in \citet{Raveri2024}. We have a minimum learning rate threshold of $10^{-5}$.
    \item Early stopping when learning no longer improves.
    \item Batch size of 128 for stable gradient updates.
    \item Shuffling of training data between epochs to prevent order-dependent learning.
\end{itemize}

\subsection{Network architecture for the 2nd moments} \label{sec:compression_moments}

For this relatively compact datavector of 120 elements, we found that a standard fully-connected neural network architecture works  well. This is likely because the datavector is both structurally simpler and dimensionally smaller than the one used for Betti numbers. For this statistic we also use separate neural networks for each target parameter and  train with an MSE loss function.

In particular we use a multi-layer perceptron with varying depths ($4-7$ hidden layers) and widths for each cosmological parameter. The network employs ReLU and Leaky ReLU activation functions and is trained with the Adam optimizer to minimize mean squared error. We optimize the network architecture (layer count, widths, activations, and learning rate) using \textsc{Optuna} \citep{optuna_2019}, a hyperparameter optimization framework. We provide more details about the specific architecture in Appendix~\ref{app:architecture}.

It is worth noting that we also explored alternative dimensionality reduction techniques, including the linear methods recently proposed by \citet{Park2025}, which were reported to outperform neural networks for parameter compression in certain contexts. However, our experiments showed that these claims do not generalize to scenarios involving wide prior ranges for the intrinsic alignment parameter ($A_{\rm IA}$). Specifically, while these methods performed adequately when $A_{\rm IA}$ was constrained to a narrow range ($-1$ to $1$), their performance deteriorated significantly when applied to our analysis with the broader prior range ($-3$ to $3$).

\subsection{Combining Betti numbers with 2nd moments}

We explored two approaches for combining Betti numbers with second moments: (1) joint compression of both statistics with a single network, and (2) separate compression followed by a concatenation of the compressed data vectors. Our experiments show that the latter approach yields significantly better results, allowing each specialized network to extract parameter-specific information without interference between the different types of statistics. Our  combination procedure works then as follows:

\begin{enumerate}
    \item We independently compress the Betti numbers using the CNN architecture and the second moments using the fully-connected architecture, both detailed in App.~\ref{app:architecture}.
    \item For each target parameter ($\Omega_\mathrm{m}$, $S_8$, $w$, and $A_\mathrm{IA}$), we obtain two compressed values: one from the topological features and one from the second moments. These compressed representations are concatenated to form an 8-dimensional  vector per simulation.
    \item The combined vector is then used as input to the neural density estimator for likelihood estimation as described in Sec.~\ref{sec:nle}.
\end{enumerate}

Our neural density estimator approach naturally accounts for any cross-correlations between the two statistics, as the NDE learns the joint likelihood $p(t|\theta)$ where $t$ now includes information from both Betti numbers and second moments. The network learns not only the individual dependencies of each statistic on the parameters, but also their joint covariance structure through the simulation data.

\subsection{Training-testing division of simulation data}

Our analysis employs 14,279  filtered simulations\footnote{We exclude simulations with $w<-1.05$ as well as a single faulty simulation that we have found doing this analysis, which is identified by the parameter combination \texttt{noiserel=0}, \texttt{mock=L}, \texttt{rot=1}, and \texttt{run=89}.}, divided into two  datasets. To divide the simulations we follow a similar approach as in \citet{Gatti2025} and \citet{Jeffrey2025}. The first dataset, comprising 60\% of simulations (8,567 simulations that correspond to noise realizations 0, 1, and 2), is used for the compression ---with 90\% allocated to training and 10\% to validation. The second dataset, containing the remaining 40\% of simulations (5,712) from two additional noise realizations, serves as our test set\footnote{While these two sets share the same underlying shear maps, they have independent noise properties and nuisance parameter values. This partial independence is sufficient for our analysis given the noise-dominated nature of the DES Y3 data. Table~\ref{tab:sims_summary} provides a  breakdown of the simulation configurations, detailing precisely which input parameters are varied across different subsets of the simulation suite.}. We keep a large test set because our complete methodology follows a two-stage process: first, we train compression models on the initial dataset; then, we apply these models to compress the second dataset, which subsequently provides inputs to the neural density estimator (NDE). Through this second step, we learn the likelihood $p(t|\theta)$ of the compressed statistics $t$ given input parameters $\theta$. We restrict the NDE training exclusively to the second set to avoid propagating overfitting effects from the compression to the final uncertainties. The posteriors are then anyway rigorously validated through a coverage test, as described in Sec.~\ref{sec:coverage_test}.

\section{Validation} \label{sec:validation}

In this section, we describe the validation tests specific to this work. We also note that our analysis relies on the extensive validation of the simulations and critical components of the SBI pipeline conducted in the companion works of \citet{Gatti2024, Gatti2025} and \citet{Jeffrey2025}.

\subsection{Impact of baryonic effects}

\begin{figure}
    \centering
    \includegraphics[width=0.45\textwidth]{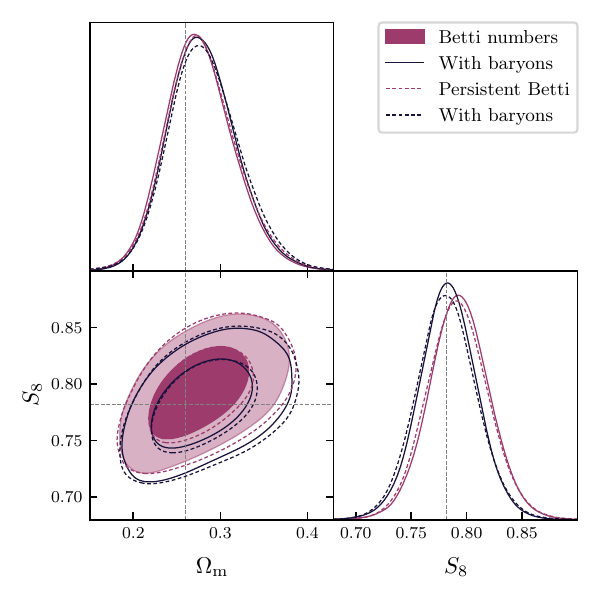}
    \caption{Impact of baryonic effects on the cosmological posteriors for Betti numbers and persistent Betti numbers. We use as input the mean of 800 \texttt{CosmoGrid} simulations with and without baryonic effects. For standard Betti numbers, the combined 2D significance of the shift in the $\Omega_\mathrm{m}$-$S_8$ plane due to baryonic effects is 0.13$\sigma$, and for persistent Betti numbers  0.12$\sigma$. The minimal shifts demonstrate that our analysis is robust to baryonic physics, allowing us to include information from all smoothing scales in our analysis.}
    \label{fig:baryons_lfi_test}
\end{figure}

\begin{figure*}
    \centering
    \includegraphics[width=0.95\textwidth]{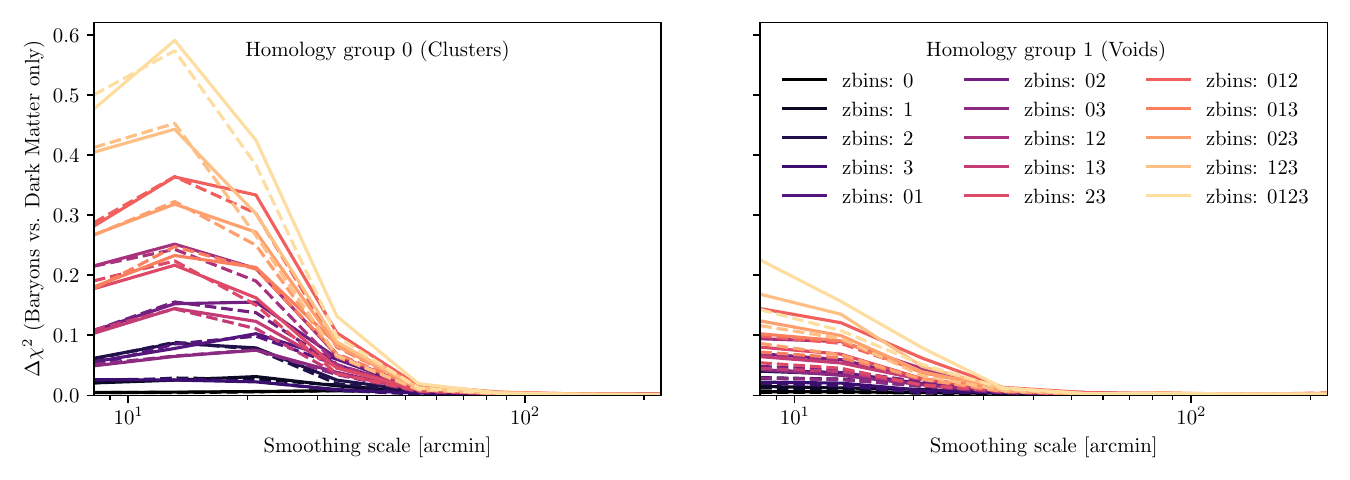}
    \caption{Impact of baryonic effects on Betti numbers across different smoothing scales and redshift bin combinations. The left panel shows results for Homology Group 0 (generally associated with clusters), while the right panel shows Homology Group 1 (generally associated with voids). Each line represents a different $z$-bin or combination of $z$-bins. Solid lines correspond to the Betti numbers and dashed lines  to persistent Betti numbers. The $\Delta\chi^2$ metric indicates the statistical difference between dark matter-only and baryonic simulations. }
    \label{fig:baryons_chi2}
\end{figure*}

We use the \texttt{CosmoGridV1} simulation suite to assess the impact of baryonic effects on our datavector. This suite has been post-processed to incorporate baryonic feedback using the baryonic correction model, as detailed in Sec.~\ref{sec:cosmogrid}. Our analysis utilizes 800 realizations of DES Y3-like convergence maps, derived from 200 independent full-sky simulations, with four DES footprints extracted from each. To quantify the baryonic impact in a way that mimics an idealized noiseless scenario, we compare the mean datavectors computed across all 800 realizations, one with baryonic effects included and one without.

Our goal is to ensure that our statistics and the scales used in this study are not significantly affected by potential baryonic contamination, which is not accounted for in the Gower Street simulations that we use for inference. Specifically, we verify that the mean of the marginalized two-dimensional posterior distribution of $\Omega_\mathrm{m}$ and $S_8$, when analyzing data with baryonic contamination, is within $0.3\sigma$ of the mean obtained from uncontaminated data. In Fig.~\ref{fig:baryons_lfi_test} we show the impact of baryonic effects on the $\Omega_\mathrm{m}$-$S_8$ joint posterior. The measured 2D shifts are small: $0.13\sigma$ for standard Betti numbers and $0.12\sigma$ for persistent Betti numbers---both  below our predefined $0.3\sigma$ threshold for validation. When combined with second moment statistics (see Fig.~\ref{fig:app:baryons-4pars}), the 2D difference is still under the threshold even if the contours shrink, with 0.15$\sigma$ for both the standard Betti numbers and the persistent Betti numbers. This implies that we do not need to remove any scales affected by baryonic physics from our analysis. 

While these results demonstrate robustness to baryonic effects, it is important to contextualize our validation approach. Baryonification methods have been primarily validated for position and angle-averaged statistics rather than topological features specifically. Still, there has been partial validation through bispectrum measurements as shown in \citet{Arico2021} and \citet{Anbajagane2024}. The bispectrum captures some angular dependencies, providing a step toward validating more complex statistics. More recently, \citet{Zhou2025} demonstrated that baryonification methods perform well for higher-order statistics including wavelets and scattering transforms, offering strong support for their applicability to diverse statistical measures beyond traditional two-point functions. Nevertheless, our findings align with previous analyses, particularly the work of \citet{Gatti2024}, which demonstrated similar insensitivity to baryonic effects at the same minimum smoothing scale (8 arcmin FWHM) using wavelet phase harmonics and moments of the convergence field. This consistency across different statistical approaches provides additional confidence in our validation.  Since all these methods capture different aspects of the same underlying density field, and traditional statistics where baryonification is well-validated show minimal baryonic impact at our chosen scales, we gain confidence that our persistent homology features are similarly robust.

Moreover, in two-point correlation function analyses, both the transformation kernel from $P(k)$ to $\xi_+$ and the one of convergence correlations use the $J_0$ Bessel function, and standard cosmic shear analyses minimally remove datapoints for $\xi_+$ due to baryonic effects \citep{Amon2022, Secco2022}. Our 8 arcmin smoothing scale is significantly larger than the scale cuts typically used in two-point analyses, providing additional reason to expect minimal baryonic contamination in our convergence mass maps. The 8 arcmin minimum smoothing scale corresponds to physical lengths of approximately 2, 3, 4, and 5 Mpc$/h$ for our four tomographic bins, respectively, which extend from just outside the 1-halo term into the 2-halo term regime, where baryonic effects are already expected to have less impact on the matter distribution.

In Fig.~\ref{fig:baryons_chi2} we also show the direct impact of baryonic contamination on each part of the datavector. We find that baryonic effects have a substantially stronger influence on Homology Group 0 (connected components/clusters) compared to Homology Group 1 (holes/voids), as might be intuitively expected since baryonic feedback primarily affects high-density regions.  The different impact is confirmed in the corresponding 2D shifts presented in Table~\ref{tab:summary_statistics}, which measure 0.12$\sigma$ for $\beta_0$ (Homology Group 0) and only 0.03$\sigma$ for $\beta_1$ (Homology Group 1). Additionally, we observe that cross-correlations between redshift bins are significantly more impacted by baryonic effects than auto-correlations. In Fig.~\ref{fig:baryons} we show that this is due to the absolute difference between the Betti numbers for baryons vs. dark matter only simulations being larger, rather than the uncertainties decreasing. 

This differential impact across homology groups and correlation types represents an advantage of persistent homology as a summary statistic. Even though our current analysis meets the validation criteria without requiring mitigation strategies, these findings suggest that baryonic effects could be effectively isolated and mitigated in future analyses by differentially weighting various components of the persistent homology data vector. This contrasts with other methods (such as map-level compressions or statistics that do not distinguish between different physical structures) where separating baryonic effects is more challenging. We believe that this physically interpretable nature of persistent homology   makes it particularly valuable for robust cosmological inference as we progress toward Stage IV surveys with increased sensitivity. For more  visualizations of the impact of baryonic effects to the Betti numbers and persistent histograms we refer the reader to Appendix~\ref{app:full_params}.

\subsection{Posterior Probability Validation: Coverage Tests}\label{sec:coverage_test}

\begin{figure}
   \centering
    \includegraphics[width=0.4\textwidth]{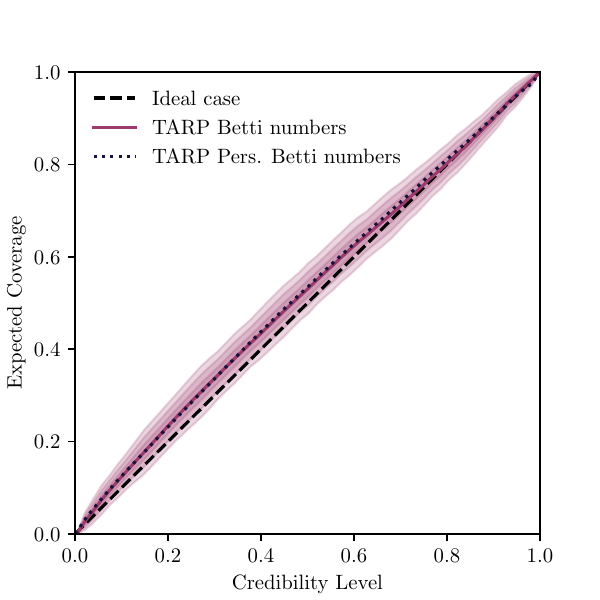}
\caption{Coverage test results for the combination of (persistent) Betti numbers and second moments, obtained using the \texttt{TARP} package. The y-axis shows the expected coverage probability --- the fraction of times that credible regions of a given confidence level actually contain the true parameter when averaged over many tests. The x-axis shows the credibility level --- the nominal confidence level of those regions (e.g., 0.68 for 68\% credible regions). The diagonal dashed line represents perfect calibration where a credibility level of X matches an expected coverage of X, which our results follow within the uncertainties (represented by the violet shaded areas, constrained by the finite number of posterior samples used).}
    \label{fig:coverage_test}
\end{figure}

We verify that the confidence levels derived from simulation-based inference are accurately estimated by performing an empirical coverage test. In this procedure, the inference process is repeated multiple times to ensure that the resulting posteriors correctly reflect the true parameter probabilities. In Bayesian inference, while there is no theoretical guarantee that credible intervals match frequentist coverage properties, empirical validation of this correspondence is valuable. Our tests demonstrate that our 68\% credible regions contain the true parameters approximately 68\% of the time, suggesting our posteriors accurately quantify parameter uncertainty.

To perform this test, we ran the inference $\sim 550$ times, each time excluding one mock data vector (selected at random) from the set used to train the neural likelihood estimator. For each excluded data vector, we computed the likelihood, and derived the corresponding posterior. We then evaluated the coverage probabilities in the four-dimensional parameter space defined by $\Omega_{\rm m}$, $S_8$, $w$ and $A_{\textrm{IA}}$, using the \texttt{TARP} package, which implements the Tests of Accuracy with Random Points (TARP) method to estimate the coverage of posterior estimators \citep{Lemos2023}. For each of the 568 posterior realizations, we drew a set of random reference points in parameter space and computed the fraction of posterior samples that lay closer to the reference point than the true parameter value. This fraction defines a credible region centered on the reference point; repeating this process allows us to estimate how often the true parameters fall within a region of given credibility (e.g., 68 percent). The TARP method guarantees both necessary and sufficient conditions for accurate coverage. Our procedure follows the methodology used in the companion analyses by \citet{Jeffrey2025} and \citet{Gatti2024}.

The results, shown in Fig.~\ref{fig:coverage_test}, demonstrate that the observed coverage  matches the expected credibility levels within statistical uncertainties, indicating that our posterior estimates are well calibrated.

\subsection{End-to-End Testing with CosmoGrid Simulations} \label{sec:e2e_test}

To further validate our inference pipeline, we conducted an end-to-end test using the \texttt{CosmoGrid} simulation suite. We created a noiseless test case by averaging the Betti numbers from 800 simulations at the fiducial cosmology, then processed this mean datavector through our complete analysis pipeline. To minimize stochastic variations in the neural density estimation (NDE) process, we trained the NDE five independent times and concatenated the resulting posterior chains. We also apply this same process to  our DES Y3 data analysis. 

Figure~\ref{fig:betti_vs_moments_comparison} shows the resulting constraints in the $\Omega_{\mathrm{m}}-S_8$ plane. The recovered parameters are all within 0.3$\sigma$ in the 2D plane. For both standard and persistent Betti numbers we find shifts of 0.08$\sigma$ with respect to the truth in the 2D plane and 0.2$\sigma$ when combined with the second moments. This confirms our methodology does not introduce large parameter biases.

\subsection{Blinding} \label{sec:blinding}
To avoid confirmation bias we implemented a blinding protocol. All methodology decisions in this work were made without looking at the cosmological results from the data: we designed our pipeline, determined all analysis choices, and implemented the full simulation-based inference framework prior to examining any cosmological parameter constraints from the actual data. We also wrote the paper and went through one round of the internal collaboration review process while still blind to the final results. This approach ensures that our methodology was not tuned to match expectations or previous results, which we believe significantly strengthens the robustness and reliability of  this work. 

As part of our unblinding procedure, we performed a sequence of basic checks: First, we verified that the DES Y3 Betti numbers datavector fell within the range of measurements from the Gower Street Simulations. Next, we confirmed that the compressed values remained within the parameter space covered by the Gower Street Simulations. We then verified that our second-moment-only results were consistent with the analogous findings reported in \citet{Gatti2025}. We then unblinded the Betti number results. Finally, we verified the compatibility between second-moment posteriors and Betti number posteriors before combining them.

\section{Cosmological results} \label{sec:cosmo-results}

\begin{figure}
   \centering
    \includegraphics[width=0.5\textwidth]{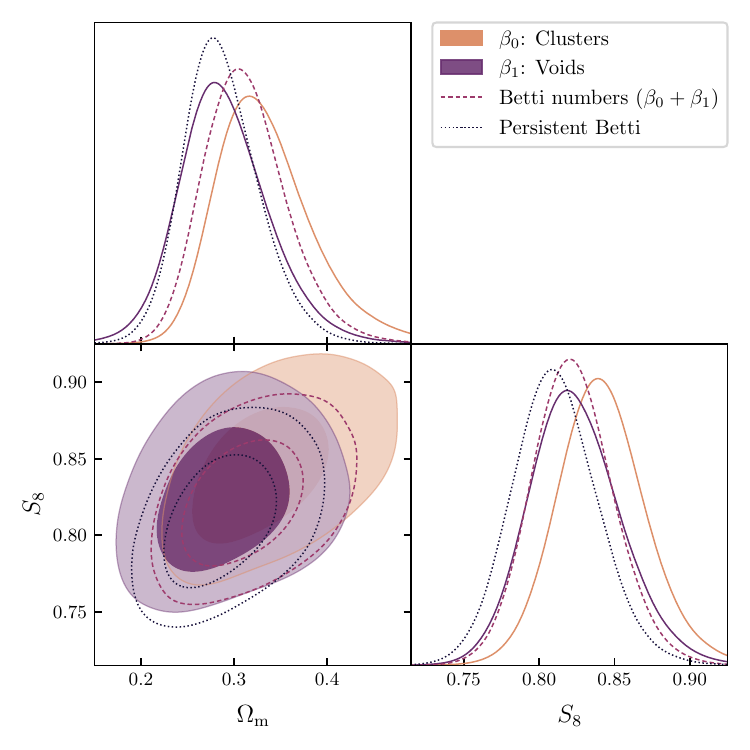}
    \caption{DES Y3 Betti numbers $w$CDM results: For the different homology groups $\beta_0$ (corresponding to overdensity regions or clusters) and $\beta_1$ (corresponding to underdensity regions or voids), along with their combination (Betti numbers) and persistent Betti numbers statistic.}
    \label{fig:desy3_cosmological_results_H0_H1}
\end{figure}

\begin{figure}
   \centering
    \includegraphics[width=0.5\textwidth]{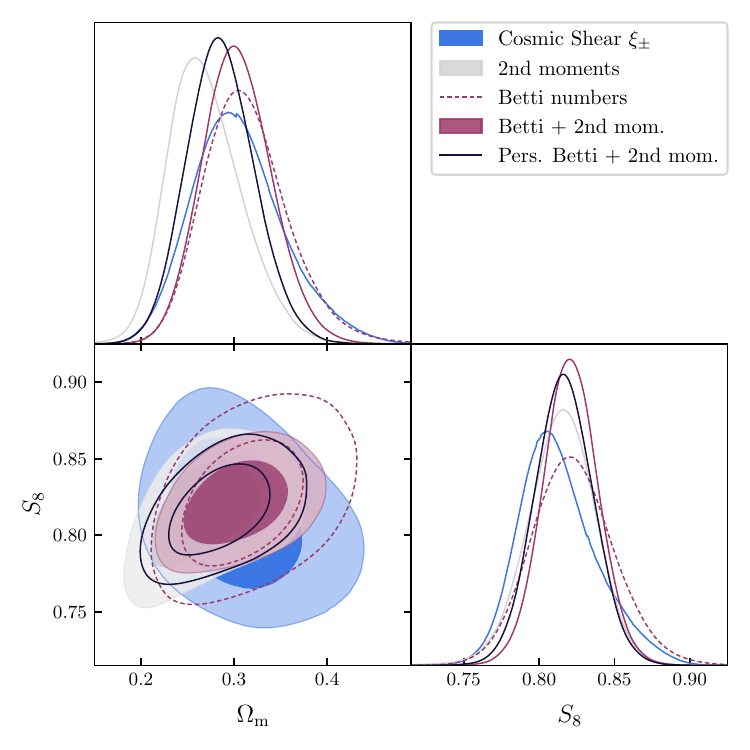}
    \caption{DES Y3 $w$CDM results: our fiducial results combine (persistent) Betti numbers with second moments. The figure of merit in the $\Omega_{\mathrm{m}}$-$S_8$ plane for each statistic is 1542 for the Betti + 2nd moments combination, which represents a 30\% improvement compared to second moments alone (1204). We also compare with the conventional cosmic shear analysis from $\xi_{\pm}$ two-point correlations from \citet{Amon2022} and \citet*{Secco2022} reanalyzed to match our priors as a reference.}
    \label{fig:desy3_cosmological_results}
\end{figure}

\begin{table*}
\centering
\caption{Results for the DES Y3 data, assuming the $w$CDM model. For each parameter we report the mean and the 68\% credible interval (except  for $w$ where we report the 68\% upper limit).}
\label{tab:parameter_constraints_des_y3}
\renewcommand{\arraystretch}{1.3} 
\begin{tabular}{lcccccc}
\toprule
\textbf{Summary Statistic} & \textbf{$S_8$} & \textbf{$\Omega_\mathrm{m}$} & \textbf{$w$} & \textbf{$A_\mathrm{IA}$} & \textbf{FoM ($\Omega_m, S_8$)} & \textbf{FoM ($\Omega_m, S_8, w, A_\mathrm{IA}$) } [$\times 10^{-3}$] \\
\midrule
$\beta_0$ (Clusters) &0.842$\pm{0.029}$ &0.333$^{+0.050}_{-0.049}$ & $< -0.76$ &0.36$^{+0.38}_{-0.37}$ & 733 & 14.6  \\
$\beta_1$ (Voids) &0.824$\pm{0.030}$ &0.290$\pm{0.047}$ & $< -0.72$  &0.72$\pm{0.46}$ & 676 & 10.2  \\
\midrule
2nd moments  & $0.814\pm{0.022}$ &0.270$^{+0.040}_{-0.039}$ & $< -0.76$ &0.40$^{+0.39}_{-0.36}$ & 1204 & 30.7 \\
Betti numbers ($\beta_0$+$\beta_1$) & 0.823$\pm{0.027}$ &0.313$\pm{0.044}$ & $< -0.69$ &0.84$\pm{0.41}$&  876 & 14.8  \\
Persistent Betti numbers &0.811$\pm{0.028}$ &0.287$\pm{0.040}$ &  <$-0.72$ &0.89$\pm{0.50}$  & 907 & 13.2  \\
Betti numbers + 2nd moments & 0.821$\pm{0.018}$ &0.304$\pm{0.037}$ &  <$-0.75$  &0.64$\pm{0.35}$ & 1542 & 42.3   \\
Persistent Betti + 2nd moments &0.817$\pm{0.019}$ &0.286$\pm{0.036}$ & <$-0.75$&0.50$^{+0.35}_{-0.34}$ & 1551 & 44.4  \\
\midrule
Cosmic shear 2pt & 0.813$^{+0.032}_{-0.022}$ & 0.303$^{+0.040}_{-0.051}$ &  $<-0.71$  & 0.34$^{+0.28}_{-0.25}$  & 901 & 40.3  \\
\bottomrule
\end{tabular}
\end{table*}

\subsection{Persistent homology DES Y3 data results}

Figure~\ref{fig:desy3_cosmological_results_H0_H1} shows the cosmological constraints from each homology group analyzed separately and then combined on DES Y3 data. Clusters ($\beta_0$) provide slightly tighter constraints than voids ($\beta_1$), with Figures of Merit (FoM\footnote{We define $\text{FoM} = \frac{1}{\sqrt{\det(\mathbf{C}_{p_1, p_2, \ldots, p_n})}}$, where $\mathbf{C}_{p_1, p_2, \ldots, p_n}$ represents the covariance matrix of parameters $p_1, p_2, \ldots, p_n$.}) in the $\Omega_m-S_8$ plane of 733 and 676, respectively. This ordering aligns with the literature, where clusters (overdensities, peaks) typically outperform voids (underdensities, local minima) in cosmological constraints. However, the performance gap is notably smaller than expected from previous studies. The mean of our \texttt{CosmoGrid} simulations show a more pronounced difference (FoM of 836 for clusters versus 718 for voids), and this gap widens substantially in the full four-dimensional parameter space, where the FoM decreases from approximately 14,600 for clusters to 10,200 for voids.

However, the relatively similar constraining power of clusters and voids in the $\Omega_m-S_8$ plane from persistent homology represents a novel finding. This reduced asymmetry, compared to traditional peak/void statistics, may reflect fundamental differences in how persistent homology extracts cosmological information. Unlike conventional methods that count features at fixed thresholds, persistent homology tracks how topological structures evolve across multiple threshold levels, potentially extracting more balanced information from both density peaks and voids. Additionally, the specific noise characteristics and systematic effects present in the DES Y3 dataset may contribute to this result.

When combined, the Betti numbers ($\beta_0$+$\beta_1$, dashed magenta) provide tighter constraints than either homology group alone, with a FoM in the $\Omega_m-S_8$ plane  of 876. The persistent Betti numbers result (dotted black) is consistent to the standard Betti numbers one, with slightly higher constraining power (FoM = 907).

Our fiducial results, presented in Figure~\ref{fig:desy3_cosmological_results}, combine Betti numbers (or persistent Betti numbers) with second moments. For the DES Y3 data realization, we find that adding Betti numbers to the second moments increases the FoM in the $\Omega_\mathrm{m}-S_8$ plane by approximately 30\% (from 1204 to 1542).

The results from persistent Betti numbers combined with second moments  are generally similar to those from standard Betti numbers, though with a slightly lower preferred value of $\Omega_\mathrm{m}$. Second moments alone  tend to prefer a lower value of $\Omega_\mathrm{m}$ compared to the Betti number statistics. The full parameter constraints are presented in Table~\ref{tab:parameter_constraints_des_y3} and also in Appendix~\ref{app:full_params_data}. Note that for all parameters that we report in such table we gain information with respect to such  analysis prior, which is displayed in Table~\ref{tab:sims_pars}.

\subsection{Comparison with other statistics}

To contextualize our results, we compare our persistent homology constraints with other summary statistics applied to the same \texttt{CosmoGrid} simulation suite and DES Y3 dataset as described in the companion papers \citet{Gatti2024, Gatti2025}, respectively, providing a direct one-to-one comparison. 

\subsubsection{Comparison on \texttt{CosmoGrid}} \label{sec:cosmogrid_comparison}

We first compare the constraining power in the ``noiseless'' case provided by the mean of \texttt{CosmoGrid} simulations. Comparing our second moments results with those from (\citealt{Gatti2024}, \textbf{\blue{G24}} here), we find a Figure of Merit (FoM) of 1053 while \blue{G24} reports 904. Our analysis yields slightly tighter $S_8$ constraints, whereas \blue{G24} achieves  better $\Omega_\mathrm{m}$ constraints.
Despite analyzing the same statistic and using the same simulations and SBI framework, we report here some methodological differences between both analyses that can explain the differences:
\begin{itemize}
\item Our analysis includes additional combinations of redshift bins beyond what \blue{G24} incorporates. Specifically, both analyses use individual redshift bins and all pairwise combinations, but we additionally incorporate every possible three-bin combination and the complete four-bin combination. Each combined map is constructed by summing the convergence  maps of the corresponding individual redshift bins, and we measure the standard deviation of these summed maps. Even though three-bin variances and the four-bin variance can be calculated from pairwise and individual terms and thus do not provide any additional cosmological information, we have tested that including these extra bins provides an additional $\simeq10$\% of constraining power for the second moments. We attribute this enhancement to the compression network's improved performance when provided with this redundant  information in the data vector.

\item We measure the standard deviation of maps after following the complete preprocessing pipeline described in Sec.~\ref{sec:processing_maps} (downgrading, smoothing, masking, and rescaling), while \blue{G24} only applies downgrading and smoothing. We note this methodological difference for completeness, though it likely contributes minimally to differences in constraining power between the analyses.
\item We optimize the compression of second moments using \textsc{Optuna} \citep{optuna_2019}, as detailed in Sec.~\ref{sec:compression_moments}, which is also contributing to improved FoM performance.
\end{itemize}

With these methodological differences established, we can meaningfully interpret the results across different statistical approaches.  The Betti numbers in our analysis (FoM = 1060) show similar constraining power to the combination of second and third moments (FoM = 1035) reported in Table V of \blue{G24}. When combined with second moments, the Betti numbers achieve a FoM of 1732, substantially exceeding the performance of second moments combined with scattering transforms (FoM = 1245) and reaching very similar power to the  combination of  second moments + third moments + scattering transforms + wavelet phase harmonics from \blue{G24} with $\mathrm{FoM} = 1733$.

This comparison suggests that persistent homology efficiently extracts non-Gaussian information at a level similar to all these other higher-order statistics combined, although the strong performance may be partly due to our CNN-based compression approach of the Betti numbers. 

\subsubsection{Comparison on DES Y3 data}\label{sec:comparisonDESY3}

\begin{figure}
   \centering
    \includegraphics[width=0.5\textwidth]{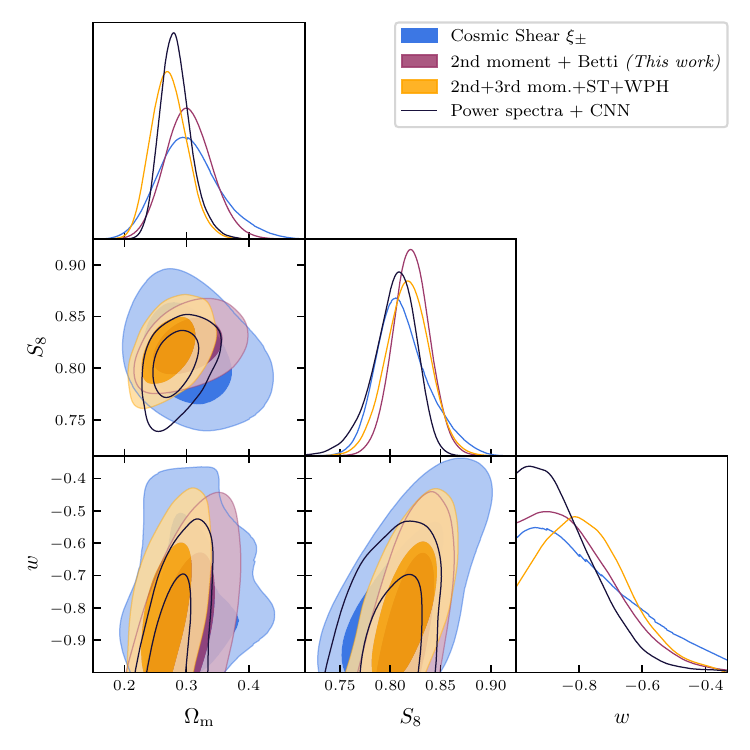}
    \caption{DES Y3 $w$CDM results: We compare the conventional cosmic shear analysis from $\xi_{\pm}$ two-point correlations from \citet{Amon2022} and \citet*{Secco2022}, reanalyzed to match our priors; the SBI analysis combining 2nd + 3rd moments + scattering transforms (ST) + wavelet harmonics (WPH) from \citet{Gatti2025} and the power spectrum + map-level CNN from \citet{Jeffrey2025}, both of which use the same framework as this work; and our fiducial result from the combination of Betti numbers and 2nd moments. Caveat: see section \ref{sec:comparisonDESY3} for $S_8$ impact of different analysis choices from \citet{Jeffrey2025}.} 
    \label{fig:desy3_results_higher_order_comparison}
\end{figure}

As shown in Fig.~\ref{fig:desy3_cosmological_results} and Table~\ref{tab:parameter_constraints_des_y3}, our fiducial results are consistent and more constraining than the ones from the cosmic shear two-point correlation function. To perform this comparison, we have reanalyzed the standard  DES Y3 $\xi_{\pm}$ results based on an analytical model. This includes the following main changes with respect to the results published in the DES Y3 cosmic shear papers from \citet{Amon2022} and \citet*{Secco2022}:

\begin{enumerate}
    \item Matching priors for the $w$CDM model ($-1 < w < -1/3$) and cosmological parameters ($n_s$, $\Omega_b$, $h$, neutrino mass) as in Table~\ref{tab:sims_pars}. You can see a comparison of the differences between our priors and the fiducial priors for the DES Y3 $w$CDM analysis in figure 2 from \citet{Gatti2024}.
    \item Using an NLA intrinsic alignment model instead of TATT, with priors matching\footnote{The IA model in the Gower Street simulations is slightly different from the one used for the DES Y3 cosmic shear re-analysis to which we compare in this work. While both are NLA models plus a clustering term (see Eqs. 2–4 in \citealt{Gatti2024}), the theory implementation used in the cosmic shear analysis estimates these contributions using tree-level perturbation theory, whereas our implementation directly uses the clustering of the simulation, which is larger at intermediate to small scales.} Table~\ref{tab:sims_pars}. 
    \item Excluding the shear-ratio likelihood from \citet*{Sanchez2022}, which primarily constrains IA and redshift parameters.
    \item Employing the \textsc{HyperRank} method \citep{Cordero2022} for redshift uncertainties, consistent with our simulation-based inference framework.
\end{enumerate}

Then, we find our combination of Betti numbers with second moments to be 70\% more constraining in the Figure of Merit in the $\Omega_\mathrm{m}-S_8$ plane compared to the reanalyzed cosmic shear two-point analysis (FoM = 1542 vs. 901). However, note that we find our second moments results alone to be already more constraining than the 2pt cosmic shear ones in the $\Omega_\mathrm{m}-S_8$ plane (FoM = 1204 vs. 901). This could be partly due to different scale weightings (see e.g. figure 20 of \citealt{Gatti2022}) and the different treatment of the intrinsic alignment source clustering term. Moreover, the statistical power appears to be distributed differently across parameters, as cosmic shear yields better constraints in the full four-dimensional parameter space with an FoM ($\Omega_m$, $S_8$, $w$, $A_\mathrm{IA}$) of $40.3\times10^{3}$ compared to $30.7\times10^{3}$ for the second moments.

We also compare our results with other higher-order statistics applied to DES Y3 data in Fig.~\ref{fig:desy3_results_higher_order_comparison}. Our constraints show very good agreement with those from \citet{Gatti2025} (\textbf{\blue{G25}}), who used 2nd + 3rd moments + scattering transforms (ST) + wavelet harmonics (WPH), and with \citet{Jeffrey2025} (\textbf{\blue{J25}}), who did a map-level CNN compression combining with the power spectrum. This comparison is particularly valuable since all analyses use the same underlying data, simulations, and SBI framework, allowing us to isolate the impact of the different statistics and compression approaches. An important consideration though is that \blue{J25} excluded the intrinsic alignment parameter $A_\mathrm{IA}$ from their compressed data vector, whereas both \blue{G25} and our analysis include this component, which has now been found to yield tighter $S_8$ constraints.

When comparing constraining power across these statistics, our Betti + 2nd moments combination achieves a Figure of Merit of 1542, which is competitive with \blue{G25} most complete combination of 2nd + 3rd moments + ST + WPH (FoM = 1725). Remarkably, our approach using only two summary statistics (Betti numbers and 2nd moments) achieves similar constraining power of this more complex combination that utilizes four distinct statistics. This may be partially attributable to methodological differences in how we analyze our statistics. As detailed in Sec.~\ref{sec:cosmogrid_comparison}, our 2nd moment implementation differs from \blue{G25} in several aspects, including the inclusion of more redshift bin correlations and an optimized compression network. Additionally, it is important to note that the constraining power of higher-order statistics can vary significantly depending on the specific noise realization of the data. \citet*{Gomes-Sugiyama2025}  demonstrate that the improvement from adding higher-order statistics to Gaussian statistics can range from 24\% to 170\% depending on the noise realization, which may contribute to the differences in FoM between our ``noiseless'' simulated results and actual noisy data constraints.

In terms of 1D parameter constraints, our Betti + 2nd moments combination ($S_8 = 0.821 \pm 0.018$) achieves similar precision on $S_8$ as \blue{G25} most complete combination ($S_8 = 0.817\pm 0.021$), with nearly identical central values. The constraints on $w$ are also similar. For $\Omega_\mathrm{m}$, our constraint ($0.304\pm0.037$) centers on a slightly higher value than \blue{G25} ($0.273\pm 0.029$), and with slightly broader uncertainty width. Similarly, when comparing with \blue{J25}'s map-level CNN technique, we find our method offers slightly tighter constraints on $S_8$ but  broader constraints on $\Omega_\mathrm{m}$ and $w$.

\section{Conclusions and Outlook} \label{sec:conclusion}

In this work, we have presented the first application of spherical persistent homology to a galaxy survey, analyzing the Dark Energy Survey Year 3 (DES Y3) weak lensing mass maps using the \textsc{TopoS2} algorithm and a robust simulation-based inference framework.

\subsection{Main results}
Our analysis demonstrates that topological methods provide a powerful framework for extracting cosmological information:
\begin{enumerate}
   \item \textbf{Consistent Cosmological Constraints:} Assuming a $w$CDM model, our combined analysis of Betti numbers and second moments yields $S_8 = 0.821 \pm 0.018$ and $\Omega_\mathrm{m} = 0.304\pm0.037$, with $w < -0.81$. These results are consistent with the DES Y3 cosmic shear analysis from \citet{Amon2022} and \citet*{Secco2022} and with other DES Y3 higher order statistics analyses from \citet{Gatti2025} and \citet{Jeffrey2025} .
   \item \textbf{Improved Statistical Power:} Our analysis demonstrates that persistent homology effectively captures non-Gaussian information beyond traditional statistics, with improvements that vary for different input realizations. For the DES Y3 data, combining Betti numbers with second moments increases the Figure of Merit in the $\Omega_\mathrm{m}-S_8$ plane by approximately 30\%, while in a noiseless simulated environment, this improvement is  70\%. Compared to cosmic shear two-point functions, our approach delivers a 70\% improvement in the same $\Omega_\mathrm{m}-S_8$ plane. When compared with other DES Y3 higher-order statistics, persistent homology is competitive, achieving similar  constraining power of the most complete combination of statistics from \citet{Gatti2025} that utilizes four  statistics (2nd + 3rd moments + scattering transforms + wavelet harmonics). When compared to \citet{Jeffrey2025}'s map-level CNN analysis, our analysis has  smaller error bars on $S_8$, whereas their results have lower uncertainty on $\Omega_\mathrm{m}$ and $w$.
   \item \textbf{Robustness to Systematics:} Validation tests using the \texttt{CosmoGrid} simulations demonstrate that our analysis is robust to baryonic effects, with parameter shifts below $0.3\sigma$ in the $\Omega_m-S_8$ plane. The differential response of topological features --- with Homology Group 0 (typically associated to clusters) showing greater sensitivity to baryonic effects than Homology Group 1 (voids) --- offers a natural path for isolating systematics in future analyses if needed.
\end{enumerate}

\subsection{Methodological advances}

We have developed the following key advantages with respect to previous work:
\begin{enumerate}
   \item \textbf{Spherical Analysis:} The implementation of spherical persistent homology enables analysis of complete surveys in their native geometry without requiring fragmentation or flat-sky projections—an approach increasingly valuable for upcoming large-area surveys.
   \item \textbf{Efficient Information Compression:} Our convolutional neural network architecture successfully compresses a complex and high-dimensional datavector while preserving cosmological information across different smoothing scales and redshift bins.
   \item \textbf{Optimized Combination Strategy:} We have found that compressing Betti numbers and second moments separately before combining them at the likelihood level yields significantly better results than joint compression. 
\end{enumerate}

\subsection{Future prospects}

This methodology opens several promising avenues for future research. The spherical persistent homology framework developed here is readily transferable to upcoming wide-field surveys such as LSST (\citealt{abell2009lsst}), Roman   \citep{Roman} and Euclid \citep{Euclid}, where its native handling of curved-sky geometry will be increasingly valuable. Future work could also explore combining this topological approach with other cosmological probes within a unified analysis framework, potentially further enhancing parameter constraints and also going beyond persistent Betti numbers. Persistence diagrams themselves contain richer structural information---such as feature lifetimes---that could be harnessed through more advanced summary statistics to further improve cosmological inference. The robustness and increased statistical power demonstrated in this work also makes persistent homology a promising tool for testing extensions to the standard cosmological model, including modified gravity scenarios and evolving dark energy. Additionally, the differential response of topological features to baryonic effects—with clusters showing greater sensitivity than voids—suggests potential for constraining feedback models directly from observations, offering a new window into the complex physics of galaxy formation. 

\section*{Contribution statement}
JP: Conceptualized and designed the project, led the data analysis, and served as primary author of the manuscript. MG: Provided the DES Y3 simulation products, contributed the Simulation-Based Inference (SBI) codebase, and offered technical support throughout the analysis phase. CD: Co-led the implementation of the Convolutional Neural Network (CNN) compression algorithms and helped with the interpretation of the persistent homology measurements. PP: Contributed the TopoS2 codebase with ongoing technical support for installation and implementation throughout the project timeline, and provided specialized expertise in interpreting the persistent homology measurements. CC: Provided feedback on the manuscript draft and contributed to discussions during the analysis phase. NJ: Co-led the generation of the Gower Street simulations, proposed the simulation-based inference workflow. LW: Co-led the generation of the Gower Street simulations. DA, SS, AT: DES internal
reviewers of the paper. MG, ES, AA, MB, MT, AC, CD,
NM, ANA, IH, DG, GB, MJ, LFS, AF, JM, RPR, RC,
CC, SP, IT, JP, JEP, CS: Production of the DES shape
catalog. JM, AA, AA, CS, SE, JD, JM, DG, GB, MT,
SD, AC, NM, BY, MR: Production of the DES redshift
distribution. SE,BY, NK, EH, YZ: Production of the
DES Balrog. MJ, GB, AA, CD, PFL, KB, IH, MG, AR:
Production of the DES PSF. Builders: The remaining
authors have made contributions to this paper that include, but are not limited to, the construction of DECam
and other aspects of collecting the data; data processing and calibration; developing broadly used methods,
codes, and simulations; running the pipelines and validation tests; and promoting the science analysis

\section*{Acknowledgments}

We would like to thank Daniel Sanz-Alonso, Sven Heydenreich and Joachim Harnois-Deraps for helpful early discussions on this work.  

JP has been supported by the Eric and Wendy Schmidt AI in Science Postdoctoral Fellowship, a Schmidt Futures program. PP is partially supported by the Simons-Ashoka fellowship (grant no: 993444). This work was supported by collaborative visits funded by the Cosmology and Astroparticle Student and Postdoc Exchange Network (CASPEN). The Gower Street simulations were generated under the DiRAC project p153 ‘Likelihood-free inference with the Dark Energy Survey’ (ACSP255/ACSC1) using DiRAC (STFC) HPC facilities (\texttt{www.dirac.ac.uk}).

Funding for the DES Projects has been provided by the U.S. Department of Energy, the U.S. National Science Foundation, the Ministry of Science and Education of Spain, 
the Science and Technology Facilities Council of the United Kingdom, the Higher Education Funding Council for England, the National Center for Supercomputing 
Applications at the University of Illinois at Urbana-Champaign, the Kavli Institute of Cosmological Physics at the University of Chicago, 
the Center for Cosmology and Astro-Particle Physics at the Ohio State University,
the Mitchell Institute for Fundamental Physics and Astronomy at Texas A\&M University, Financiadora de Estudos e Projetos, 
Funda{\c c}{\~a}o Carlos Chagas Filho de Amparo {\`a} Pesquisa do Estado do Rio de Janeiro, Conselho Nacional de Desenvolvimento Cient{\'i}fico e Tecnol{\'o}gico and 
the Minist{\'e}rio da Ci{\^e}ncia, Tecnologia e Inova{\c c}{\~a}o, the Deutsche Forschungsgemeinschaft and the Collaborating Institutions in the Dark Energy Survey. 

The Collaborating Institutions are Argonne National Laboratory, the University of California at Santa Cruz, the University of Cambridge, Centro de Investigaciones Energ{\'e}ticas, 
Medioambientales y Tecnol{\'o}gicas-Madrid, the University of Chicago, University College London, the DES-Brazil Consortium, the University of Edinburgh, 
the Eidgen{\"o}ssische Technische Hochschule (ETH) Z{\"u}rich, 
Fermi National Accelerator Laboratory, the University of Illinois at Urbana-Champaign, the Institut de Ci{\`e}ncies de l'Espai (IEEC/CSIC), 
the Institut de F{\'i}sica d'Altes Energies, Lawrence Berkeley National Laboratory, the Ludwig-Maximilians Universit{\"a}t M{\"u}nchen and the associated Excellence Cluster Universe, 
the University of Michigan, NSF NOIRLab, the University of Nottingham, The Ohio State University, the University of Pennsylvania, the University of Portsmouth, 
SLAC National Accelerator Laboratory, Stanford University, the University of Sussex, Texas A\&M University, and the OzDES Membership Consortium.

Based in part on observations at NSF Cerro Tololo Inter-American Observatory at NSF NOIRLab (NOIRLab Prop. ID 2012B-0001; PI: J. Frieman), which is managed by the Association of Universities for Research in Astronomy (AURA) under a cooperative agreement with the National Science Foundation.

The DES data management system is supported by the National Science Foundation under Grant Numbers AST-1138766 and AST-1536171.
The DES participants from Spanish institutions are partially supported by MICINN under grants PID2021-123012, PID2021-128989 PID2022-141079, SEV-2016-0588, CEX2020-001058-M and CEX2020-001007-S, some of which include ERDF funds from the European Union. IFAE is partially funded by the CERCA program of the Generalitat de Catalunya.

We  acknowledge support from the Brazilian Instituto Nacional de Ci\^encia
e Tecnologia (INCT) do e-Universo (CNPq grant 465376/2014-2).

This document was prepared by the DES Collaboration using the resources of the Fermi National Accelerator Laboratory (Fermilab), a U.S. Department of Energy, Office of Science, Office of High Energy Physics HEP User Facility. Fermilab is managed by Fermi Forward Discovery Group, LLC, acting under Contract No. 89243024CSC000002.

\section*{Data Availability}
The DES Y3 Weak Lensing maps can be found here: \texttt{https://des.ncsa.illinois.edu/releases/y3a2/Y3massmaps}. The Gower Street Simulation Suite can be found here: \texttt{http://www.star.ucl.ac.uk/GowerStreetSims/README.html}.

\bibliography{library}
\bibliographystyle{mnras_2author}

\section*{Affiliations}
\begin{small}
$^{1}$ Nordita, KTH Royal Institute of Technology and Stockholm University, Hannes Alfv\'ens v\"ag 12, SE-10691 Stockholm, Sweden\\
$^{2}$ Oskar Klein Centre, Department of Physics, Stockholm University, SE-106 91 Stockholm, Sweden\\
$^{3}$ Department of Astronomy and Astrophysics, University of Chicago, Chicago, IL 60637, USA\\
$^{4}$ Kavli Institute for Cosmological Physics, University of Chicago, Chicago, IL 60637, USA\\
$^{5}$ Department of Physics and Astronomy, University of Pennsylvania, Philadelphia, PA 19104, USA\\
$^{6}$ Universit\'e Grenoble Alpes, CNRS, LPSC-IN2P3, 38000 Grenoble, France\\
$^{7}$ Bennett University, Plot no 8-12, Greater Noida, UP, 201310, India\\
$^{8}$ Ashoka University, National Capital Region P.O., Plot no. 2, Rajiv Gandhi Education City, Rai, Sonipat, Haryana, India\\
$^{9}$ Department of Physics \& Astronomy, University College London, Gower Street, London, WC1E 6BT, UK\\
$^{10}$ Department of Physics, ETH Zurich, Wolfgang-Pauli-Strasse 16, CH-8093 Zurich, Switzerland\\
$^{11}$ Institute of Space Sciences (ICE, CSIC),  Campus UAB, Carrer de Can Magrans, s/n,  08193 Barcelona, Spain\\
$^{12}$ Department of Astrophysical Sciences, Princeton University, Peyton Hall, Princeton, NJ 08544, USA\\
$^{13}$ Physics Department, 2320 Chamberlin Hall, University of Wisconsin-Madison, 1150 University Avenue Madison, WI  53706-1390\\
$^{14}$ Department of Physics, Carnegie Mellon University, Pittsburgh, Pennsylvania 15312, USA\\
$^{15}$ NSF AI Planning Institute for Physics of the Future, Carnegie Mellon University, Pittsburgh, PA 15213, USA\\
$^{16}$ Department of Physics, Duke University Durham, NC 27708, USA\\
$^{17}$ NASA Goddard Space Flight Center, 8800 Greenbelt Rd, Greenbelt, MD 20771, USA\\
$^{18}$ Kavli Institute for Particle Astrophysics \& Cosmology, P. O. Box 2450, Stanford University, Stanford, CA 94305, USA\\
$^{19}$ Lawrence Berkeley National Laboratory, 1 Cyclotron Road, Berkeley, CA 94720, USA\\
$^{20}$ Fermi National Accelerator Laboratory, P. O. Box 500, Batavia, IL 60510, USA\\
$^{21}$ Department of Physics and Astronomy, University of Waterloo, 200 University Ave W, Waterloo, ON N2L 3G1, Canada\\
$^{22}$ California Institute of Technology, 1200 East California Blvd, MC 249-17, Pasadena, CA 91125, USA\\
$^{23}$ SLAC National Accelerator Laboratory, Menlo Park, CA 94025, USA\\
$^{24}$ University Observatory, Faculty of Physics, Ludwig-Maximilians-Universit\"at, Scheinerstr. 1, 81679 Munich, Germany\\
$^{25}$ Jet Propulsion Laboratory, California Institute of Technology, 4800 Oak Grove Dr., Pasadena, CA 91109, USA\\
$^{26}$ School of Physics and Astronomy, Cardiff University, CF24 3AA, UK\\
$^{27}$ Department of Applied Mathematics and Theoretical Physics, University of Cambridge, Cambridge CB3 0WA, UK\\
$^{28}$ Instituto de F\'isica Gleb Wataghin, Universidade Estadual de Campinas, 13083-859, Campinas, SP, Brazil\\
$^{29}$ Department of Physics, University of Genova and INFN, Via Dodecaneso 33, 16146, Genova, Italy\\
$^{30}$ Jodrell Bank Center for Astrophysics, School of Physics and Astronomy, University of Manchester, Oxford Road, Manchester, M13 9PL, UK\\
$^{31}$ Brookhaven National Laboratory, Bldg 510, Upton, NY 11973, USA\\
$^{32}$ Department of Physics and Astronomy, Stony Brook University, Stony Brook, NY 11794, USA\\
$^{33}$ Institut de Recherche en Astrophysique et Plan\'etologie (IRAP), Universit\'e de Toulouse, CNRS, UPS, CNES, 14 Av. Edouard Belin, 31400 Toulouse, France\\
$^{34}$ Excellence Cluster Origins, Boltzmannstr.\ 2, 85748 Garching, Germany\\
$^{35}$ Max Planck Institute for Extraterrestrial Physics, Giessenbachstrasse, 85748 Garching, Germany\\
$^{36}$ Universit\"ats-Sternwarte, Fakult\"at f\"ur Physik, Ludwig-Maximilians Universit\"at M\"unchen, Scheinerstr. 1, 81679 M\"unchen, Germany\\
$^{37}$ Cerro Tololo Inter-American Observatory, NSF's National Optical-Infrared Astronomy Research Laboratory, Casilla 603, La Serena, Chile\\
$^{38}$ Institute for Astronomy, University of Edinburgh, Edinburgh EH9 3HJ, UK\\
$^{39}$ INAF-Osservatorio Astronomico di Trieste, via G. B. Tiepolo 11, I-34143 Trieste, Italy\\
$^{40}$ Laborat\'orio Interinstitucional de e-Astronomia - LIneA, Av. Pastor Martin Luther King Jr, 126 Del Castilho, Nova Am\'erica Offices, Torre 3000/sala 817 CEP: 20765-000, Brazil\\
$^{41}$ Physik-Institut, University of Zürich, Winterthurerstrasse 190, CH-8057 Zürich, Switzerland\\
$^{42}$ Department of Physics, Northeastern University, Boston, MA 02115, USA\\
$^{43}$ Institut de F\'{\i}sica d'Altes Energies (IFAE), The Barcelona Institute of Science and Technology, Campus UAB, 08193 Bellaterra (Barcelona) Spain\\
$^{44}$ Instituto de Astrofisica de Canarias, E-38205 La Laguna, Tenerife, Spain\\
$^{45}$ Universidad de La Laguna, Dpto. Astrofísica, E-38206 La Laguna, Tenerife, Spain\\
$^{46}$ Physics Department, William Jewell College, Liberty, MO, 64068\\
$^{47}$ Centro de Investigaciones Energ\'eticas, Medioambientales y Tecnol\'ogicas (CIEMAT), Madrid, Spain\\
$^{48}$ Department of Physics, IIT Hyderabad, Kandi, Telangana 502285, India\\
$^{49}$ Hamburger Sternwarte, Universit\"{a}t Hamburg, Gojenbergsweg 112, 21029 Hamburg, Germany\\
$^{50}$ Instituto de Fisica Teorica UAM/CSIC, Universidad Autonoma de Madrid, 28049 Madrid, Spain\\
$^{51}$ Center for Astrophysical Surveys, National Center for Supercomputing Applications, 1205 West Clark St., Urbana, IL 61801, USA\\
$^{52}$ Department of Astronomy, University of Illinois at Urbana-Champaign, 1002 W. Green Street, Urbana, IL 61801, USA\\
$^{53}$ School of Mathematics and Physics, University of Queensland,  Brisbane, QLD 4072, Australia\\
$^{54}$ Santa Cruz Institute for Particle Physics, Santa Cruz, CA 95064, USA\\
$^{55}$ Center for Cosmology and Astro-Particle Physics, The Ohio State University, Columbus, OH 43210, USA\\
$^{56}$ Department of Physics, The Ohio State University, Columbus, OH 43210, USA\\
$^{57}$ Center for Astrophysics $\vert$ Harvard \& Smithsonian, 60 Garden Street, Cambridge, MA 02138, USA\\
$^{58}$ Australian Astronomical Optics, Macquarie University, North Ryde, NSW 2113, Australia\\
$^{59}$ Lowell Observatory, 1400 Mars Hill Rd, Flagstaff, AZ 86001, USA\\
$^{60}$ George P. and Cynthia Woods Mitchell Institute for Fundamental Physics and Astronomy, and Department of Physics and Astronomy, Texas A\&M University, College Station, TX 77843,  USA\\
$^{61}$ Instituci\'o Catalana de Recerca i Estudis Avan\c{c}ats, E-08010 Barcelona, Spain\\
$^{62}$ Observat\'orio Nacional, Rua Gal. Jos\'e Cristino 77, Rio de Janeiro, RJ - 20921-400, Brazil\\
$^{63}$ Centro de Tecnologia da Informa\c{c}\~ao Renato Archer, Amarais, Campinas, SP 13069-901\\
$^{64}$ Ruhr University Bochum, Faculty of Physics and Astronomy, Astronomical Institute, German Centre for Cosmological Lensing, 44780 Bochum, Germany\\
$^{65}$ Instituto de F\'\i sica, UFRGS, Caixa Postal 15051, Porto Alegre, RS - 91501-970, Brazil\\
$^{66}$ Physics Department, Lancaster University, Lancaster, LA1 4YB, UK\\
$^{67}$ Computer Science and Mathematics Division, Oak Ridge National Laboratory, Oak Ridge, TN 37831\\
$^{68}$ Institute of Cosmology and Gravitation, University of Portsmouth, Portsmouth, PO1 3FX, UK\\
$^{70}$ Department of Astronomy, University of California, Berkeley,  501 Campbell Hall, Berkeley, CA 94720, USA\\
\end{small}

\appendix

\section{Compression Architectures} \label{app:architecture}

\subsection{Architecture for the Betti numbers}

The network first reshapes the input to a tensor with shape (40, 8, 30) to better capture the relationship between different smoothing scales (8), redshift bins (15) and homology groups (2). Thus, we have 30 channels that correspond to the various tomographic bins and homology groups. The \texttt{SeparableConv2D}\footnote{\texttt{keras.io/api/layers/convolution\_layers/}} layers are the backbone of our architecture, consisting of two sequential operations: a depthwise convolution that processes each input channel independently, followed by a pointwise convolution that combines information across channels. This approach significantly reduces the parameter count compared to standard convolutions and allows the network to initially learn features specific to each redshift bin configuration and topological structure (clusters or voids) separately. We order the channels by redshift bin combinations with both homology groups ($H_i$) adjacent (e.g., $z_1-H_0$, $z_1-H_1$, $z_2-H_0$, $z_2-H_1$). We have tested  ordering  grouping all $H_0$ channels (clusters) together followed by all $H_1$ (holes) and found that it performs slightly worse\footnote{The ordering of channels primarily matters during the pointwise convolution stage of the SeparableConv2D layers, where adjacent channels are combined. This ordering becomes less important after the standard Conv2D layers, which process all channels simultaneously.}.

We list the convolutions and pooling operations in Table~\ref{tab:network_architecture} and describe them here:
\begin{enumerate}
    \item An initial depthwise separable convolution with 32 filters and a (3, 1) kernel: This kernel examines 3 consecutive points along each Betti curve without mixing smoothing scales, allowing the network to detect localized topological features. 
    \item A larger separable convolution with a (5, 3) kernel to identify broader patterns, followed by average pooling to reduce dimensionality. The (5, 3) kernel examines both more points on each Betti curve (5) and spans across multiple adjacent smoothing scales (3), enabling the network to detect patterns that depend on both the shape of individual curves and how these shapes vary across different physical scales.
    \item Two standard convolutional layers with decreasing filter counts (16 and 8) to combine features across all channels. 
    \item A fully connected output section with dropout (30\%) for regularization and robustness against overfitting.
\end{enumerate}

The network uses the inverse hyperbolic sine (asinh) activation function throughout, which effectively handles both small and large values while maintaining gradient flow across different scales. This activation function provides advantages over ReLU for our data, which contains both positive and negative values after normalization. We use a batch size of 128 and 300 epochs. Here we provide the code to implement such network:

\begin{lstlisting}[style=shadedcode, language=Python]
def create_2d_parameter_model(input_shape, activation=tf.math.asinh):
    def reshape_input(x):
        # Reshape from (batch, 40, 240) to (batch, 40, 8, 30)
        return tf.reshape(x, (-1, 40, 8, 30))
    
    model = Sequential([
        tfkl.Lambda(reshape_input, input_shape=input_shape),
        
        # Initial depthwise separable conv
        tfkl.SeparableConv2D(32, (3, 1), activation=activation, padding='same'),
        tfkl.BatchNormalization(),
        
        # Larger kernel for broader patterns
        tfkl.SeparableConv2D(32, (5, 3), activation=activation, padding='same'),
        tfkl.BatchNormalization(),
        tfkl.AveragePooling2D((2, 1)),
        
        # Regular convs for feature combination
        tfkl.Conv2D(16, (3, 3), activation=activation, padding='same'),
        tfkl.BatchNormalization(),
        tfkl.AveragePooling2D((2, 2)),
        
        tfkl.Conv2D(8, (3, 3), activation=activation, padding='same'),
        
        tfkl.Flatten(),
        tfkl.Dropout(0.3),
        tfkl.Dense(32, activation=activation),
        tfkl.Dense(1, activation=None),
    ])
    
    model.compile(optimizer=Adam(learning_rate=1e-3), loss="mse")
    return model
\end{lstlisting}

\begin{table}
\caption{Summary of the neural network architecture (a CNN) used for compressing the Betti numbers. The network first reshapes the input to exploit the natural structure of the data across different scales and redshift bins.}
\centering
\begin{tabular}{|c|c|c|c|}
\hline
\textbf{Layer Type} & \textbf{Output Shape} & \textbf{Kernel Size} & \textbf{Parameters} \\
\hline
Input & (40, 240) & - & 0 \\
Reshape & (40, 8, 30) & - & 0 \\
SeparableConv2D & (40, 8, 32) & (3, 1) & 1,082 \\
BatchNormalization & (40, 8, 32) & - & 128 \\
SeparableConv2D & (40, 8, 32) & (5, 3) & 1,536 \\
BatchNormalization & (40, 8, 32) & - & 128 \\
AveragePooling2D & (20, 8, 32) & (2, 1) & 0 \\
Conv2D & (20, 8, 16) & (3, 3) & 4,624 \\
BatchNormalization & (20, 8, 16) & - & 64 \\
AveragePooling2D & (10, 4, 16) & (2, 2) & 0 \\
Conv2D & (10, 4, 8) & (3, 3) & 1,160 \\
Flatten & (320) & - & 0 \\
Dropout (0.3) & (320) & - & 0 \\
Dense & (32) & - & 10,272 \\
Dense & (1) & - & 33 \\
\hline
\multicolumn{3}{|r|}{\textbf{Total Parameters:}} & 19,027 \\
\hline
\end{tabular}
\label{tab:network_architecture}
\end{table}

\subsection{Architecture of the 2nd moments}

In Table~\ref{tab:second_moments_architecture} we provide the details of the fully connected neural network that we use to compress the 2nd moments.

\begin{table}
\caption{Summary of the (fully connected) neural network architecture used for compressing the 2nd moments.}
\centering
\begin{tabular}{|c|c|c|c|}
\hline
\textbf{Layer Type} & \textbf{Output Shape} & \textbf{Activation} & \textbf{Parameters} \\
\hline
Input & (120) & - & 0 \\
Dense & (258) & - & 31,218 \\
LeakyReLU & (258) & $\alpha=0.3$ & 0 \\
Dense & (1033) & - & 267,547 \\
ReLU & (1033) & - & 0 \\
Dense & (281) & - & 290,554 \\
ReLU & (281) & - & 0 \\
Dense & (703) & - & 198,246 \\
LeakyReLU & (703) & $\alpha=0.3$ & 0 \\
Dense & (364) & - & 256,256 \\
LeakyReLU & (364) & $\alpha=0.3$ & 0 \\
Dense & (112) & - & 40,880 \\
ReLU & (112) & - & 0 \\
Dense & (1479) & - & 167,127 \\
LeakyReLU & (1479) & $\alpha=0.3$ & 0 \\
Dense & (1) & - & 1,480 \\
\hline
\multicolumn{3}{|r|}{\textbf{Total Parameters:}} & 1,253,308 \\
\hline
\end{tabular}
\label{tab:second_moments_architecture}
\end{table}

\section{Full parameter posteriors} \label{app:full_params_data}
\begin{figure}
    \centering
    \includegraphics[width=0.48\textwidth]{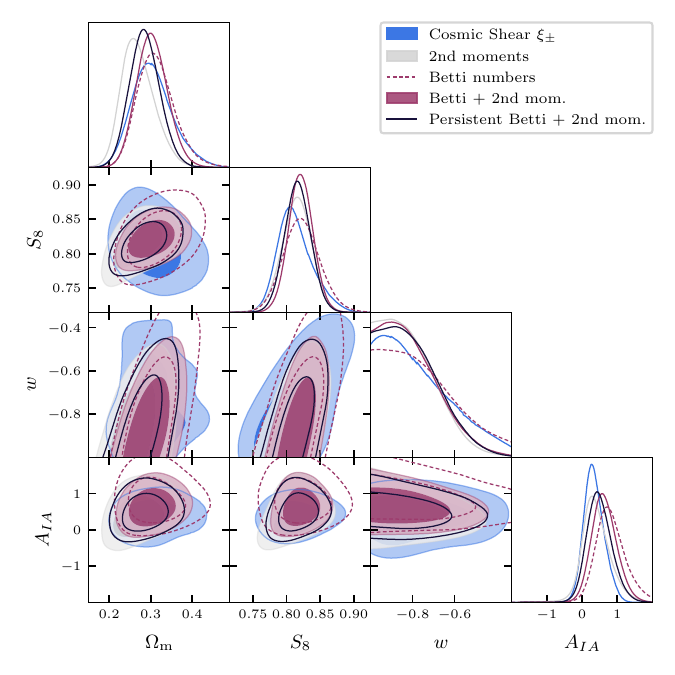}
    \caption{DES Y3 results: Full parameter constraints in the four-dimensional space ($\Omega_{\mathrm{m}}$, $S_8$, $w$, and $A_\mathrm{IA}$) using different summary statistics. }
\label{fig:app:data-4pars}
\end{figure}

In Fig.~\ref{fig:app:data-4pars} we show the full parameter posterior for DES Y3 data. We find that the combination of second moments and Betti numbers is synergetic in all the parameter space, not only in the $\Omega_\mathrm{m}-S_8$ plane.

\section{Detailed analysis of baryonic effects on topological statistics}
\label{app:full_params}

\begin{figure}
    \centering
    \includegraphics[width=0.48\textwidth]{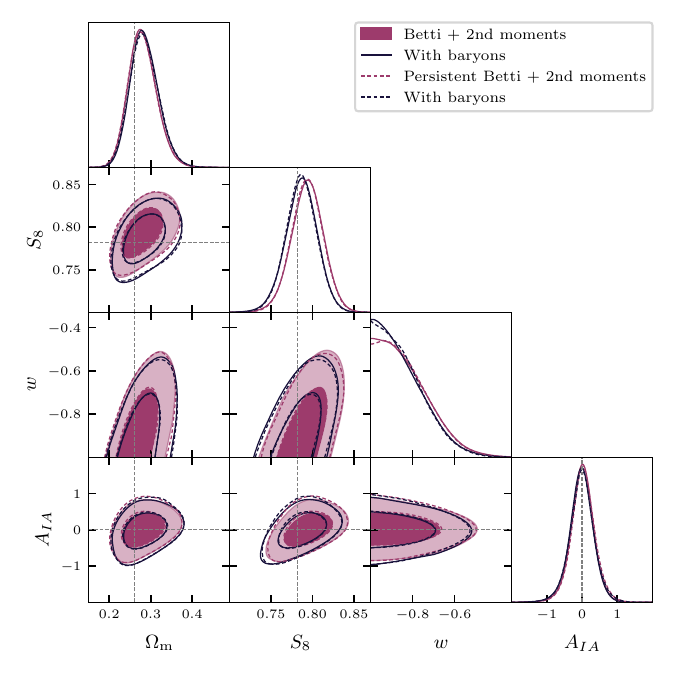}
    \caption{Impact of baryonic effects in the full parameter space constraints for the combined Betti numbers + 2nd moments analysis. Solid lines show the standard Betti numbers (purple) and persistent Betti numbers (black) for dark matter-only simulations, while dashed lines show the corresponding results including baryonic effects. }
    \label{fig:app:baryons-4pars}
\end{figure}

In the main text, we focused primarily on the $\Omega_\mathrm{m}$-$S_8$ plane when presenting the impact of baryonic effects on cosmological constraints derived from Betti numbers and persistent homology. Here, we present the full parameter space that includes the dark energy equation of state parameter $w$ and the intrinsic alignment amplitude $A_\mathrm{IA}$, shown in Fig.~\ref{fig:app:baryons-4pars} for both standard and persistent Betti numbers combined with second-moment statistics. We also present the 1D shifts for each parameter in Table~\ref{tab:summary_statistics}, on top of the already discussed 2D shifts.

\begin{figure*}
   \centering
   \includegraphics[width=\textwidth]{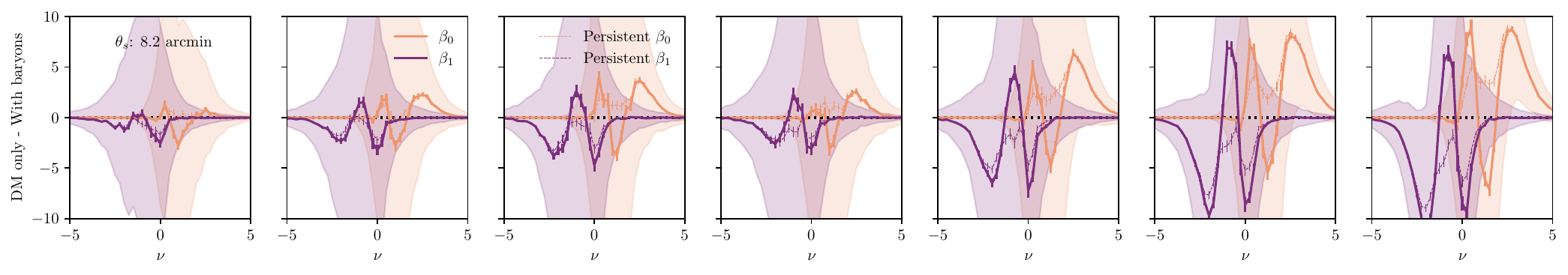}
\includegraphics[width=\textwidth]{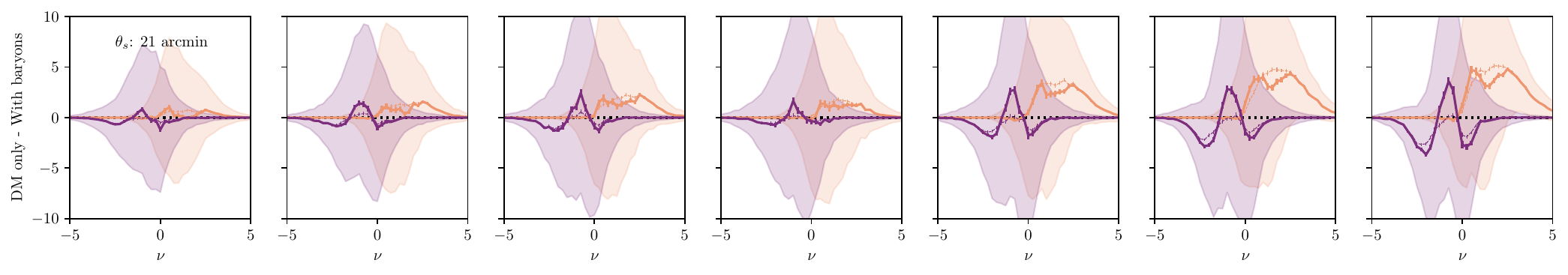}
    \includegraphics[width=\textwidth]{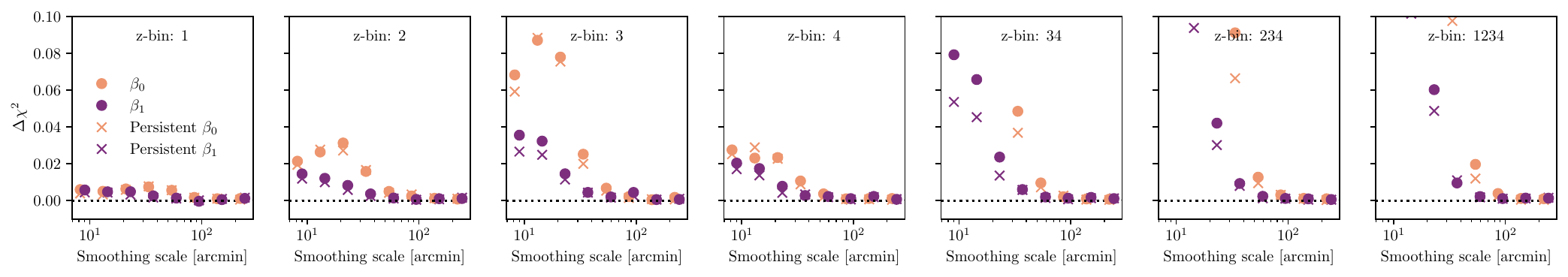}
    \caption{Impact of baryonic effects to the Betti numbers, for the clusters ($\beta_0$) and the voids ($\beta_1$). The above two rows show the difference between a contaminated Betti numbers datavector and the fiducial one, for the mean of 800 \texttt{CosmoGrid} simulations for two different smoothing scales. We show the error on the mean as errorbars on each point and the standard deviation of the sample as the shaded regions (which would correspond to the uncertainty for a single realization and is what effectively goes into the $\Delta \chi^2$ computation below). The bottom row shows the statistical significance for each of the smoothing scales and redshift bins. }
    \label{fig:baryons}
\end{figure*}

Figure~\ref{fig:baryons} illustrates how baryonic effects modify the Betti numbers for the smallest smoothing scale (8.2 arcmin, top row) and for a larger scale (21 arcmin, middle row). The bottom row quantifies these differences using a $\Delta\chi^2$ metric across all smoothing scales and redshift bins. As described in the main text, homology group 0 (connected components/clusters) shows larger deviations than homology group 1 (holes/voids), confirming that baryonic feedback predominantly affects high-density regions. The impact varies across redshift bins, with the third bin exhibiting the strongest contamination --- likely reflecting a trade-off between lensing efficiency (larger at higher redshift) and baryonic effects (stronger at lower redshift).

\begin{figure*}
   \centering
  \includegraphics[width=\textwidth]{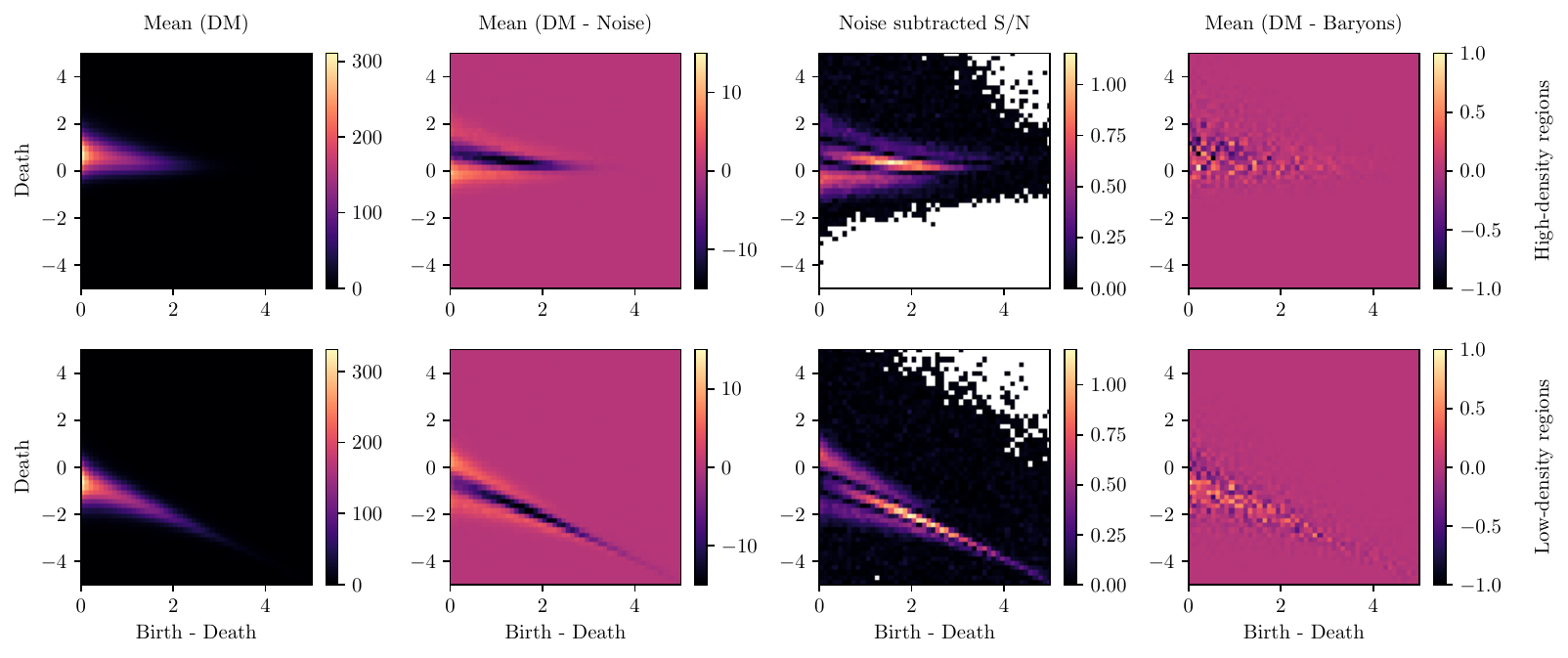}
   \caption{\texttt{CosmoGrid} 2D histograms of the persistence diagrams for the third redshift bin and smallest smoothing scale. The $x$-axis of the diagrams displays the difference between the birth and death of each event, also referred to as persistence. In the first column we show the mean of the 800 dark matter (DM) only simulations i.e. $\left<\mathrm{DM}_i \right>$, on the second one the difference between the DM simulations and the shape noise only ones $\left<\mathrm{DM}_i -\mathrm{Noise}_i\right>$, on the third one the signal-to-noise (S/N), defined as $\left< \mathrm{DM}_i-\mathrm{Noise}_i \right>/\sigma(\mathrm{DM}_i)$ and in the last column the difference with respect to simulations that include baryonic effects.}
   \label{fig:2d_histograms}
\end{figure*}

Figure~\ref{fig:2d_histograms} provides the equivalent plot for 2D histograms of persistence diagrams, for the third redshift bin and the smallest smoothing scale, which is the combination (within the autocorrelation bins) that displays the strongest effects.

\begin{table}
\centering
\caption{Summary of shifts in cosmological parameters due to baryonic effects across for the different data combinations. All shifts are reported in units of standard deviations ($\sigma$). The 1D shifts are computed between the means of the marginalized posterior distributions for each parameter. The 2D shift refers to the difference between the peaks of the posterior distributions in the $\Omega_\mathrm{m}$-$S_8$ plane.}
\label{tab:summary_statistics}
\begin{tabular}{lcccc}
\toprule
\textbf{Summary Statistic} & \textbf{$\Delta\Omega_\mathrm{m}$} & \textbf{$\Delta S_8$} & 2D shift \\
 & ($\sigma$) & ($\sigma$) & ($\sigma$)   \\
 \midrule
$\beta_0$ (Clusters) & 0.11 & 0.31 & 0.12 \\
$\beta_1$ (Voids) & 0.02 & 0.21 & 0.03 \\
\midrule 
Betti numbers $(\beta_0 + \beta_1)$ & 0.09 & 0.37   & 0.13  \\
Persistent Betti numbers & 0.09 & 0.36 & 0.12   \\
Betti numbers + 2nd moments & 0.13 & 0.36 & 0.15   \\
Persistent Betti numbers + 2nd moments & 0.12 & 0.37 &  0.15  \\
\bottomrule
\end{tabular}
\end{table}

\bsp	
\label{lastpage}
\end{document}